\documentclass[iop,revtex4]{emulateapj}
\usepackage{natbib}
\usepackage{amsfonts}
\usepackage{amssymb}
\usepackage{amsmath,amsthm}
\usepackage{enumerate}
\usepackage{amsmath, amssymb, mathrsfs}
\usepackage{rotate}
\usepackage{lscape}
\bibpunct{(}{)}{;}{a}{}{,}
\def\teff{T$_{\rm eff}$}

\def\logg{$\log$~{\it g}}
\def\vt{v$_{\rm t}$}

\def\kms{km s$^{-1}$}
\def\feh{$\rm[Fe/H]$}
\def\cd24{CD~$-24^{\circ}$17504}

\slugcomment{Accepted to The Astrophysical Journal 15 June 2015}
\shorttitle{A new abundance analysis of \cd24}
\shortauthors{Jacobson \& Frebel}

\begin{document}
\title{\cd24\ revisited: a new comprehensive element abundance analysis\altaffilmark{*}}

\author{
Heather R.\ Jacobson\altaffilmark{1},
Anna Frebel\altaffilmark{1}
}

 \altaffiltext{*}{This work is based on data obtained from the European
   Southern Observatory (ESO) Science Archive Facility and 
   associated with Programs 68.D-0094(A) and 073.D-0024(A).  This
   work is also based on data obtained from the Keck Observatory
   Archive (KOA), which is operated by the W.M. Keck Obsevatory and
   the NASA Exoplanet Science Institute (NExScI), under contract with
   the National Aeronautics and Space Administration.  These data are
   associated with Program C01H (P.I. M\'{e}lendez).}

\altaffiltext{1}{Kavli Institute for Astrophysics and Space Research
  and Department of Physics, Massachusetts Institute of Technology, 77
  Massachusetts Avenue, Cambridge, MA 02139, USA}

\begin{abstract}
With $\rm [Fe/H]\sim-3.3$, \cd24\ is a canonical metal-poor main
sequence turn-off star.  Though it has appeared in numerous literature
studies, the most comprehensive abundance analysis for the star based
on high resolution, high signal-to-noise spectra is nearly 15 years
old.  
We present a new detailed abundance analysis for 21 elements
based on combined archival Keck-HIRES and VLT-UVES spectra of the star
that is higher in both spectral resolution and signal-to-noise than
previous data.  Our
results for many elements are very similar to those of an earlier
comprehensive study of the star,
but we present for the first time a carbon abundance 
from the CH G-band feature as well as
improved upper limits for neutron-capture species such as Y, Ba and
Eu.  In particular, we find that \cd24\ has \feh\ $= -$3.41, 
[C/Fe] = $+$1.10, [Sr/H] = $-$4.68 and
[Ba/H] $\leq -$4.46, making it a carbon enhanced metal-poor star with
neutron-capture element abundances among the lowest measured in Milky
Way halo stars.
\end{abstract}

\section{Introduction}\label{sec:intro}
Metal-poor stars, especially those with $\rm [Fe/H] \lesssim -3$, are
highly sought after because of the information they provide about
early generations of star formation and chemical evolution in the
universe.  The number of stars known to have $\rm [Fe/H] \lesssim -3$ has
greatly expanded in recent years due to dedicated searches for such
objects 
and now is of order 10$^{3}$ \citep{BPSII,hes4,cayrel2004,lai2008,caffau_2011gto,norris13_I,aoki2013,cohen2013,roederer_313stars,fnaraa}.

With a visual magnitude bright enough (V$\sim$12) to place it in the
1892 Cordoba Durchmusterung (CD; \citealt{CDcat}) catalog, \cd24 also
appeared in catalogs of high proper motion stars, such as the 
New Luyten Catalogue of Stars with Proper Motions Larger than Two
Tenths of an Arcsecond (NLTT)
\citep{nltt80}.  In a survey for subdwarfs, \citet{ryan89} found it
in the NLTT catalog, and followup spectroscopic
studies at low and high resolution showed it to be extremely
metal-poor \citep{Ryanetal:1991,ryanSDIV91}.  The first high
resolution spectroscopic study of this star was done by
\citet{ryanSDIV91},
 and this study was superceded by a work with better data
in \citet{Norrisetal:2001}.  It remains one of the most well-studied
extremely metal-poor turn-off stars to this day, due its bright visual
magnitude.  In addition to the comprehensive element abundances
presented by Norris et al., the abundances of light elements 
 \citep{primas,melendez2004,aoki2009li,fabbian2009,hosford09,rich09},
 $\alpha$-elements \citep{israelian2001,arnone,fabbian2009,ishigaki2012}, and Fe-peak
 elements \citep{bihain2004,nissen2007,ishigaki2013} in \cd24\ have been
studied in some detail by several different authors.

Given the relatively high effective temperatures and surface gravities
of metal-poor stars near the main-sequence turn-off (MSTO), their spectra can contain few absorption
features suitable for detailed high resolution spectroscopic study.
Although \cd24\ is one of the brightest metal-poor dwarf stars known,
only upper limits for the neutron-capture species Ba and Eu can be determined.
Because \cd24\ is a
canonical metal-poor star, it is worthwhile to beat down the upper
limits to some abundance measurements to better constrain its
nucleosynthetic origin.  Spectra of resolution and signal-to-noise superior to
that used by \citet{Norrisetal:2001} (hereafter NRB01) have
since been obtained for \cd24.  Of these, only \citet{ishigaki2010,ishigaki2012,ishigaki2013} have presented
abundances for selected $\alpha$-, Fe-peak, and neutron-capture species for \cd24\ as
part of their comprehensive study of stars in the outer Milky Way
halo\footnote{We note that \citet{yong13_II} presented a
  comprehensive abundance analysis of \cd24; however, their analysis
  used the equivalent widths of NRB01, and therefore can be considered
an ``update'' of that work in an effort to place it on a homogeneous
scale with their larger sample.}.  
However, their analyses included only a few lines per species,
and do not include an evaluation of C, Li or neutron-capture species
beyond Sr and Eu.

Therefore, we consider it time to revisit \cd24\ in its own right,
separate from any large sample studies and in order to obtain as much
abundance information as possible from the best available data. 
We have searched archival databases for spectra of \cd24 and
present here the results of a comprehensive detailed abundance
analysis, with emphasis on elements previously undetected in
\cd24\ and those with upper limits.  We describe the data in 
Section~\ref{data}, the methods of our analysis in Section~\ref{analysis},
and the results in Section~\ref{results}.  A summary and conclusions
are given in Section~\ref{conc}.

\section{Archival spectra}\label{data}
The data used in this work are high resolution spectra from the ESO
and Keck archive facilities.  
Details of the individual exposures are given in
  Table~\ref{tab_spec} for all the data used in this analysis.
From the ESO science archives, we
downloaded UVES spectra of
\cd24 in the BLU437 and RED580 setups (see wavelength ranges,
  resolving power, and signal-to-noise (S/N) information in Table~\ref{tab_spec}). 
The spectra were obtained in Advanced Data Product format as part of
ESO's phase 3
infrastructure\footnote{http://www.eso.org/sci/observing/phase3.html}.
As such, they were reduced with version 5.1.5 of the
UVES pipeline and packaged as binary fits files.
Fully pipeline-reduced data of \cd24 
obtained with HIRES on the Keck telescope were
similarly obtained from the Keck Observatory Archive (KOA), 
also in the form of binary fits files.

Although archival data given in Table~\ref{tab_spec} 
vary by a factor of two in spectral
resolution, we chose to maximize
signal-to-noise ratio at the expense of spectral resolution for this
analysis in order to obtain more meaningful upper limits for elements
lacking absorption features such as Ba and Eu.
These archival data were combined in the following way.  Working with
each individual exposure, dispersion and flux information for each
echelle order were extracted and continuum-normalized using the
analysis package ``Spectroscopy Made Hard'' (SMH; \citealt{SMHref}).
A low-order cubic spline was used for normalization.  
Individual orders were then stitched together to create a single
continuous one-dimensional (1D) spectrum.  Each 1D spectrum was
then radial-velocity corrected by cross-correlation against a
normalized, rest-frame spectrum of HD~140283 and shifted to rest-frame
by scaling the wavelengths without interpolation or rebinning.

A linear wavelength scale was generated with a pixel size set equal
to the smallest pixel size of the data, ranging from the
shortest to the longest wavelengths shown in Table~\ref{tab_spec}.
A sparse matrix of size ($N_{\rm pixels,rebinned}$, $N_{\rm pixels,exposure}$)
was then created for each spectrum with a varying Gaussian kernel
along the diagonal to convolve the spectral resolution of each
exposure to that of the final, rebinned spectrum (R=51,700).  The
kernel values in each column of the matrix were normalized to sum to
1, such that multiplication of each 1D spectrum by this matrix
produced a rebinned, convolved, rest-frame spectrum while ensuring no
flux information was lost.  The rebinned spectra were then 
combined with each spectrum weighted by its variance.

The Keck HIRES
  ccd3 spectra, which span $\lambda$$\sim$7000$-$8350\AA, were
continuum-normalized, radial-velocity corrected and 
  combined separately within SMH and inspected for the presence of the
  oxygen triplet at $\lambda$7770 \AA.  No oxygen absorption features were
  visible, so this spectrum was not analyzed 
further.  Instead, we make use of O measures in the literature in our analysis. 

As NRB01 is the work we will most closely compare our results to, it
is worthwhile evaluating this new composite spectrum of \cd24 in terms
of their figure of merit, defined as $F = (R[S/N])/\lambda$, where R
is spectral resolution, S/N is signal-to-noise ratio, and $\lambda$ is
wavelength.  Their
echelle spectrum had F = 830, nearly a factor of two higher than
previous works (NRB01; their Table 1).  Here, adopting the
nominal R = 51,700, F $\approx$4400 at 4300\AA.  This value is a
factor of five higher than NRB01's value.  Details of the NRB01
spectrum are also given in Table~\ref{tab_spec} for comparison.

\section{Analysis}\label{analysis}
This section describes the details of our analysis of the composite
spectrum.

\subsection{Equivalenth width measurements}\label{ews}
For this work, we used the line list compiled in \citet{roederer10}.
Equivalent widths of all lines in the line list detected in the
spectrum were
measured by fitting Gaussians to them in an automatic fashion within SMH.  
These measures were then checked
by eye and remeasured by hand where necessary.  For the most part,
line measurement uncertainties due to errors in continuum
normalization or line blending were minimal.
 Lines for which element abundances deviated from those of other
  lines of the same species by more than 2$\sigma$ were discarded in
  the abundance analysis (Section~\ref{stelpars}).  The line measures used in the abundance
  analysis are given in Table~\ref{tab_ews}. Lines with equivalent
 widths (EWs) as small as 1.5 m\AA\ were distinguishable from the
 continuum.  
Some absorption features in Table~\ref{tab_ews} 
were evaluated using spectrum synthesis (see next section) and are
likewise indicated.

\begin{figure}[ht!]
\begin{center}
   \includegraphics[clip=true,width=8cm]{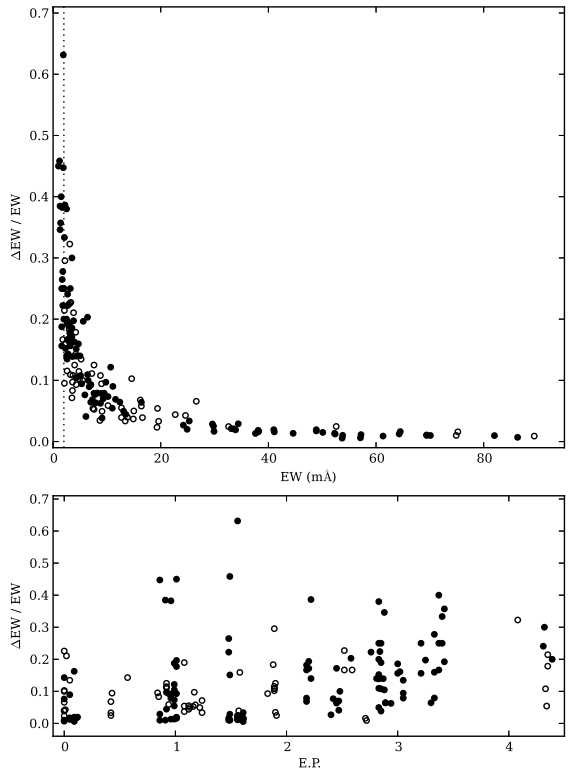} 
     \figcaption{ \label{ew_unc}
     The fraction EW uncertainty ($\Delta$EW/EW) as a function of line
     strength (top panel) and line E.P. (bottom panel).  Fe~I and
     Fe~II lines are indicated by black circles, and non-Fe species
     are given as open circles.  To guide the eye, the dotted line in
     the top panel indicates a line strength of 2 m\AA.}
 \end{center}
\end{figure}

\begin{figure*}[!ht]
\begin{center}
   \includegraphics[clip=true,width=14cm]{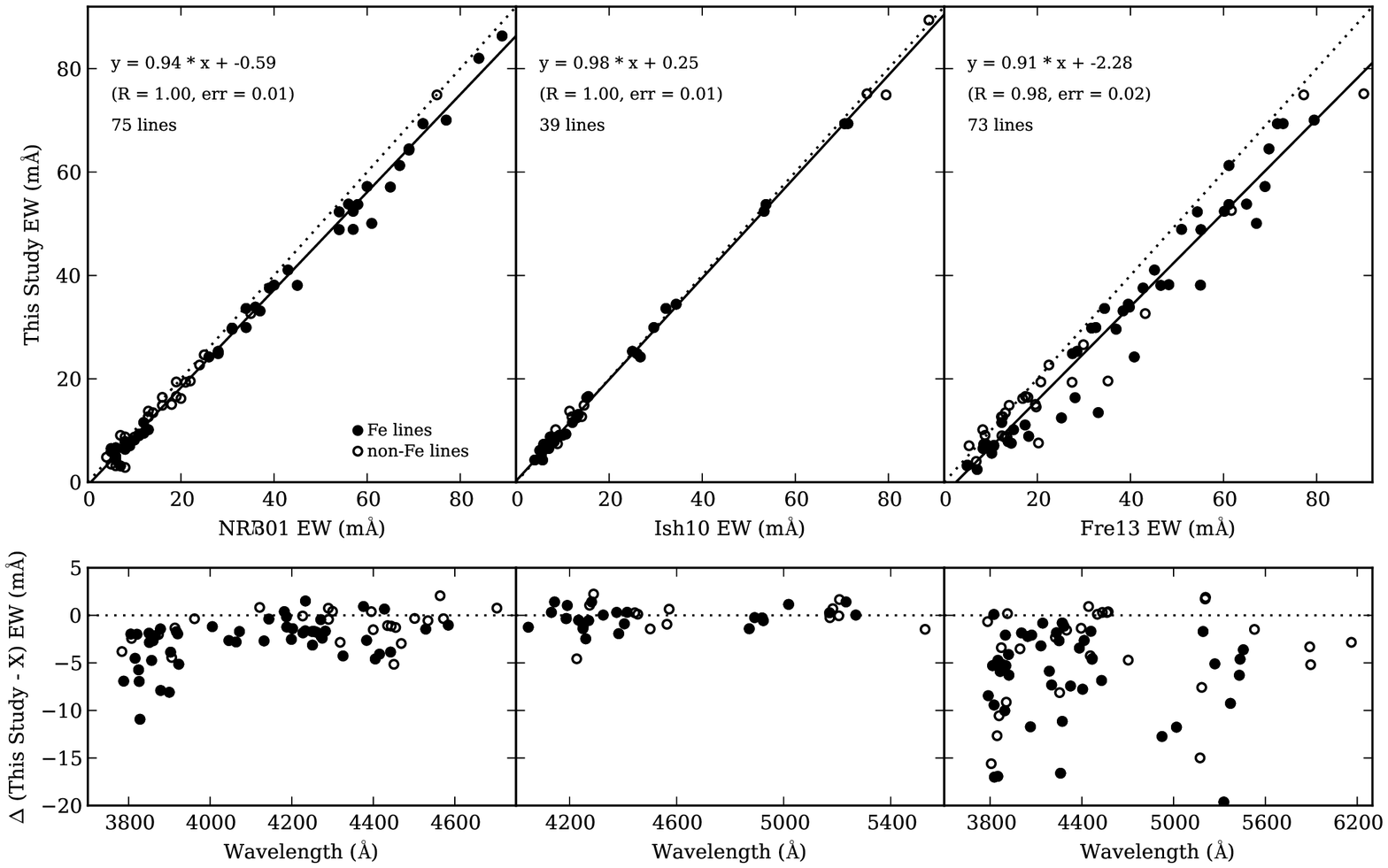} 
     \figcaption{ \label{f_ewcomp}
     Comparison of EWs measured in this study to those in the literature.  In
     the top panels, dotted lines indicate a 1:1 correlation, while
     solid lines are lines of best fit through the data.  Fe I and Fe
     II lines are indicated by filled circles.  The bottom panels show
     the difference between EW measures as a function of wavelength.}
 \end{center}
\end{figure*}

Lines with very small EW's ($<$5 mA) are more susceptible to
  errors in continuum placement than larger lines.  Such 
  errors can impact the determination of stellar parameters and
  element abundances.  To better understand this, we measured 
 minimum and maximum EWs for each line corresponding to the minimum
 and maximum values of the continuum.  Half the difference between
 these is taken as the measurement uncertainty ($\Delta$EW), which is
 also given in Table~\ref{tab_ews} for each line.  
$\Delta$EW ranges from 0.3 to 2.0 m\AA, with a mean of 0.6 m\AA\
 ($\sigma$=0.2 m\AA).
Figure~\ref{ew_unc} plots the quantity $\Delta$EW/EW versus EW and
line excitation potential (E.P.).  As can be seen, most lines have an
uncertainty of less then 20\%, but the smallest lines can have
uncertainties as large as 63\%.  The bottom panel of
Figure~\ref{ew_unc} shows the potential impact of measurement
uncertainties on determination of stellar effective temperature, as
the Fe~I lines (solid circles) show a slight trend of increasing EW
uncertainty with increasing E.P.  We explore this in more detail in Section~\ref{stelpars}.

Figure~\ref{f_ewcomp} shows our EW measures compared
to those of three studies from the literature for lines in common:
NRB01, \citet{ishigaki2010,ishigaki2012,ishigaki2013}\footnote{Although the
  results of \citet{ishigaki2012,ishigaki2013} appear separate from
  \citet{ishigaki2010}, they all use the same Subaru HDS spectrum of
  \cd24.  Therefore, we consider the EW measures from all three
  studies together.} and \citet{teff_calib}.  Stated again, the
figures of merit for the spectra in this work and in NRB01 are 4400
and 830, respectively.  The Subaru HDS spectrum used in
\citet{ishigaki2010,ishigaki2012,ishigaki2013} has F$\approx$2360,
while the Magellan-MIKE spectrum of \citet{teff_calib} has
F$\approx$490.  Characteristics of these spectra are also given in
Table~\ref{tab_spec} for comparison.

We have a total of 75 lines in common with the line list of NRB01
(left panels of Figure~\ref{f_ewcomp}). 
Our EW measures are generally smaller than theirs (by $\sim$2 m\AA), but agreement is
good for the weaker lines.  For 17
lines, the difference between our measures and those of NRB01 is 5
m\AA\ or more.  The bottom left panel of Figure~\ref{f_ewcomp} shows that
12 of 
these (predominantly Fe~I) lines are located in the region of
3750-3950\AA.  This portion of the spectrum is dominated by very
strong Balmer absorption lines, which in a star of this temperature
have very extended wings.  Visual inspection of the lines with the
largest measurement differences confirm that the majority of them are
located in the wings of these strong absorption lines, indicating that
the measurement differences could be due to differences in the
continuum normalization.

The remaining five lines with large measurement differences were not
located near any strong absorption features.  In each case, we could
not reduce the EW measurement difference by adjustment of the
continuum; to make the lines as strong as measured by NRB01 required
measuring the line well above the location of the continuum.  We
therefore attribute the measurement differences to S/N differences.

The right panels of Figure~\ref{f_ewcomp} compare our EW measures with
those measured in a MIKE spectrum of \cd24 from our earlier work
\citep{teff_calib}.  
Here, the
agreement is less good, with a larger mean offset (5.4 m\AA)
and substantially larger scatter.  We inspected each
line for which the EW difference was larger than 3 m\AA\ (45 lines) in
the MIKE spectrum from \citet{teff_calib}, which has both lower S/N  and
lower resolution  than in this work (Table~\ref{tab_spec}).  Only nine of these lie near
strong Balmer features.  For 14 lines, we found that the
EW values were consistent with each other considering the S/N of the
region; that is, that slight adjustments of the continuum level within
its uncertainty got the EW measures to agree.  For 15 further lines,
similar adjustments decreased the EW discrepency by 50\% or more.  For
some remaining lines, EW discrepencies could not be decreased. 

Lastly, the middle panels compare our measures to those in
\citet{ishigaki2010,ishigaki2012,ishigaki2013}.  
Though we only have 39 lines in common, this study is
in a sense most similar to ours in that their analysis was based on
data superior in resolution (R$\sim$55,000) and of high S/N
($\sim$250; Table~\ref{tab_spec}) .  As can be seen, agreement is excellent, with a mean
offset of $-$0.2 m\AA.  In summary, the differences in EW measures among
these different studies is representative of the varying quality of
the data they came from.

\subsection{Determination of Stellar Parameters}\label{stelpars}
For this work, we make use of the Castelli-Kurucz grid of 1D
plane-parallel model atmospheres \citep{castelli_kurucz} with
no-overshoot and the LTE analysis code MOOG 
 (May 2011 version, \citealt{moog}) that
includes treatment of Rayleigh scattering \citep{sobeck11}.
Stellar parameters for \cd24\ were determined via classical
spectroscopic techniques which use the EWs of Fe\,I and Fe\,II lines
described in the previous section.  Effective temperature was
determined by reducing any trend of Fe\,I line abundance with
excitation potential (E.P.), and microturbulent velocity was adjusted
to remove Fe\,I line abundance trends with reduced EW.  Surface
gravity, \logg, was adjusted until average Fe\,I and Fe\,II agreed
within 0.05 dex.  The
metallicity of the model atmosphere ($\rm [M/H]$) was also adjusted as needed.
This process was iterated upon until all three requirements were
satisfied, and then we applied the empirical correction to
\teff\ described in \citet{teff_calib}.    
The resulting stellar parameters are \teff\ = 6228 K,
\logg\ = 3.90, \vt\ = 1.25 km s$^{-1}$ and \feh\ = $-$3.41 ($\sigma$=0.10)
dex (Table~\ref{tab_params}).

Based on an analysis of a MIKE spectrum of \cd24 described earlier, we
found \teff\ and \vt\ values in good agreement with those found here:
6259 K and 1.40 \kms, respectively \citep{teff_calib}.  However, the
surface gravity in that work was 0.25 dex lower (\logg = 3.65), and
\feh\ was 0.18 dex higher ($-$3.23).  In \citet{teff_calib}, the Fe\,II
abundance was based on measures of two lines in \cd24, the EWs of both
being $\sim$50\% larger than found in this study (see previous
section).  To investigate the matter, we adjusted the measures of these two lines in their MIKE
spectrum 
within
comfortable limits of the noise level and repeated the stellar
parameter determination, resulting in parameters \teff=6259 K,
\logg=4.35, \vt=1.2 \kms, and \feh=$-$3.22.  This 0.7 dex adjustment
to surface gravity illustrates the
necessity of having several well-measured Fe\,II lines for
spectroscopic stellar parameter determination.  The $\sim$0.2 dex
higher metallicity compared to that found in this study can likewise
be attributed to systematically larger EWs.

\begin{figure}
\begin{center}
   \includegraphics[clip=true,width=8cm]{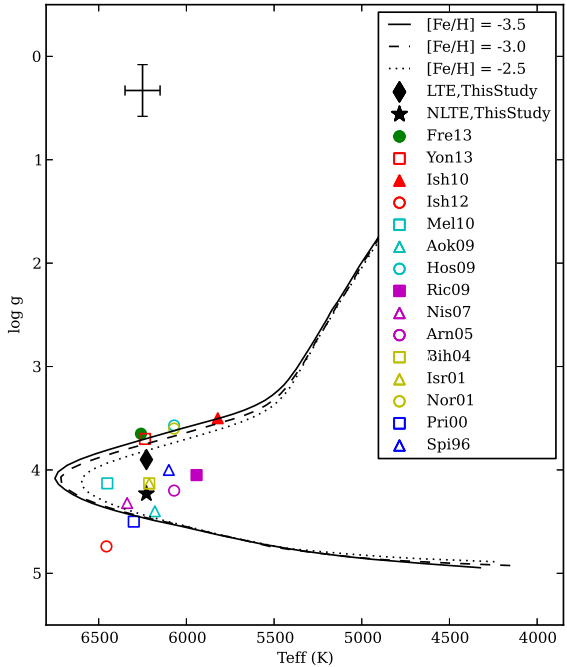} 
     \figcaption{ \label{f_hrd}
     The location of \cd24\ in the Hertzsprung-Russell diagram using
     stellar parameters from different studies.  For reference, 12 Gyr
      Yale-Yonseii isochrones with [$\alpha$/Fe] = $+0.40$ and 
     $\rm [Fe/H] = -2.5, -3.0$ and $-3.5$ are shown as dotted, dashed
      and solid lines, respectively \citep{Y2_iso}.
     Filled symbols
     indicate that stellar parameters were determined
     spectroscopically; open symbols indicate other methods were
     used.  The parameters found in this work are indicated by a
     filled diamond (LTE) and star (``NLTE''), respectively. See text for
     details.  Note that the literature results assume LTE; references
     are given in Table~\ref{tab_params}.}
 \end{center}
\end{figure}

Numerous stellar parameter determinations for \cd24\ based on
different techniques can be found in the literature.
Figure~\ref{f_hrd} shows the position of \cd24\ in the
Hertzsprung-Russell diagram using stellar parameters from various
studies, which are also listed in Table~\ref{tab_params}.  
Filled symbols represent studies that determined stellar
parameters spectroscopically, similar to that described above.  Open
symbols indicate studies that determined \teff\ via photometry and
color-temperature calibrations, or via fitting the wings of Balmer
lines.  In these cases, \logg\ was either determined via matching to
theoretical isochrones or by ionization balance of Fe\,I and Fe\,II
lines.  Figure~\ref{f_hrd} and Table~\ref{tab_params} clearly show the range of stellar
parameters these same few methods, used by different authors,
provide.  Indeed, \cd24\ can be classified as either a main sequence
dwarf star or a subgiant.

It is also well established that the assumption of local thermodynamic equilibrium
(LTE) can introduce
systematic offsets into a classical spectroscopic analysis, and that
these systematics increase with decreasing stellar metallicity and
decreasing \logg\ (e.g., \citealt{thevenin&idiart1999,asplund_araa,lind12}).  To
mitigate these effects, we determined stellar parameters for
\cd24\ following a method described in \citet{ruchti2013} and starting
with our empirically-calibrated spectroscopic \teff\ (6228
K)\footnote{Recall that this calibration places
  spectroscopically-determined \teff\ values on a rough
  ``photometric'' scale, and that this correction increases with
  decreasing \teff\ (and decreasing \logg) -- similar to the direction
  that NLTE-LTE differences increase.  Therefore this empirical
  calibration ``softens the blow'' of using purely spectroscopic
  techniques.}.  We then determined $\Delta\rm[Fe/H]$ (NLTE$-$LTE) for
Fe\,I lines in our line list that were present in the INSPECT
database\footnote{http://www.inspect-stars.com}, adopting \teff=6228 K, \logg
= 4.0, \feh=$-$3.4 and \vt=1.25 \kms\ \citep{bergemann12, lind12}.  For
the 18 Fe\,I lines, the average $\Delta\rm[Fe/H]$ = $+$0.12 dex ($\sigma$=0.04).
Therefore, $\rm[Fe\,I/H]_{LTE}$ = $-$3.41 corresponds to $\rm[Fe\,I/H]_{NLTE}$ =
$-$3.29.  Next, \logg\ was adjusted to achieve $\rm[Fe\,II/H]$ =
$-$3.29, and \vt\ was adjusted to remove any trends of Fe\,I 
abundance with line strength.  The resulting ``NLTE'' stellar
parameters are: \teff = 6228 K, \logg = 4.23, \vt = 1.00 \kms, and
\feh = $-$3.29 (Table~\ref{tab_params}).

\begin{figure*}[!t]
\begin{center}
   \includegraphics[clip=true,width=16cm]{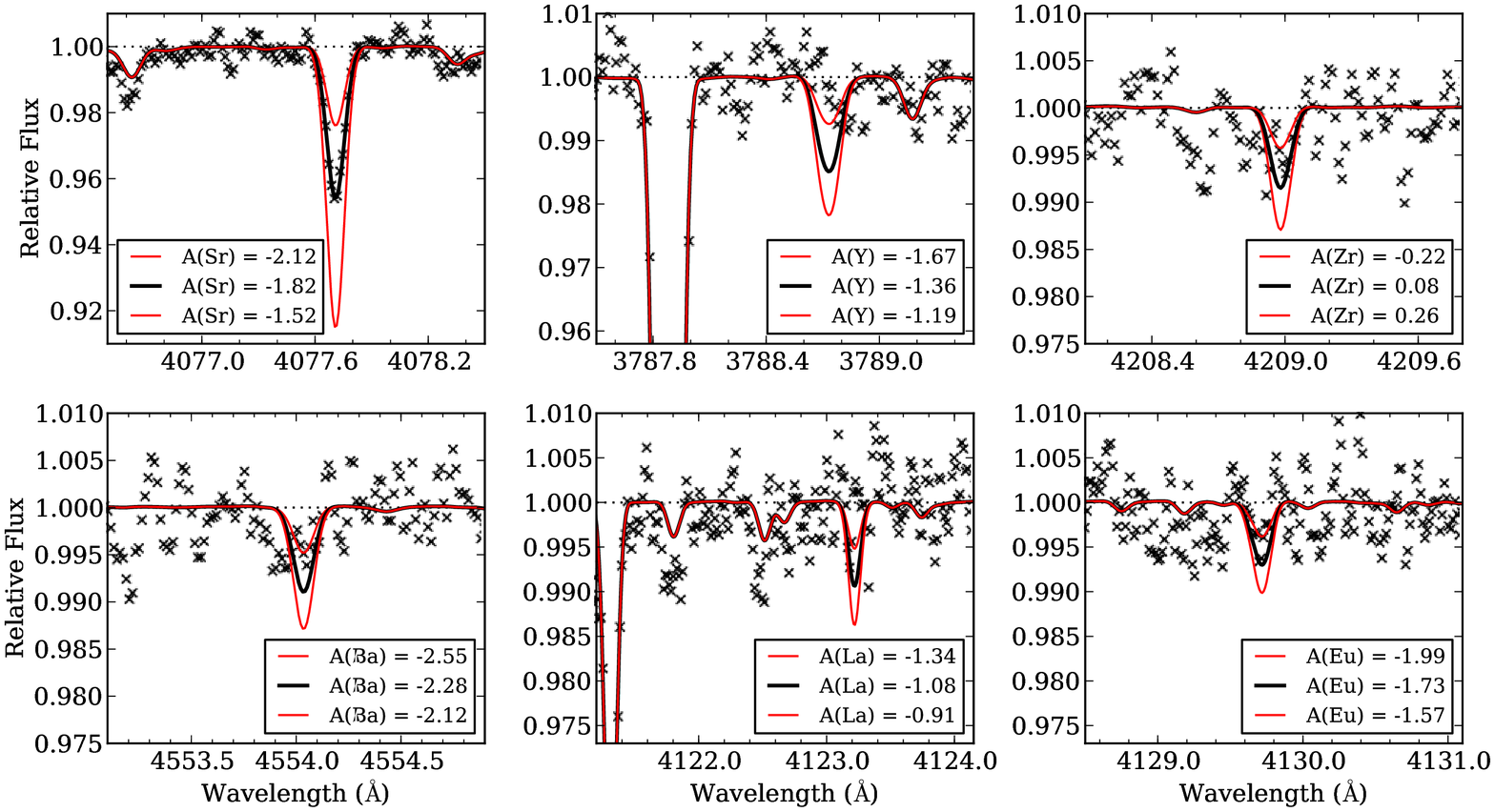} 
     \figcaption{ \label{f_synth}
     Portions of the spectrum of \cd24 at the locations of various
     neutron-capture element absorption features.  The only element
     detected, Sr, was analyzed with spectrum synthesis.  The best-fit
synthetic spectrum is indicated by a black solid line, with spectra
illustrating log$\epsilon$(Sr) $\equiv$ A(Sr) $\pm$ 0.3 dex are shown as red
     lines.  For the other elements, synthetic spectra with element
     abundances determined adopting an EW upper limit of 1 m\AA\ are
     given by black lines, with red lines representing synthetic
     spectra with abundances determined using EWs of 0.5 and 1.5 m\AA.}
 \end{center}
\end{figure*}

The LTE and ``NLTE'' parameters determined here for \cd24\ are
indicated by a black diamond and black star in Figure~\ref{f_hrd},
respectively.  As can be seen, \cd24\ appears to be a dwarf star or a
subgiant star depending on the assumption of LTE for Fe\,I lines in
the determination of \logg\ via ionization balance.  Our ``NLTE''
parameters are more consistent with those found photometrically, the
majority of which indicate \cd24\ is a dwarf star with \logg$>$4.  For
the rest of this paper, we will mostly focus on the LTE
parameters and subsequent element abundances for comparison to
literature values (which by and large assume LTE), but we will include
abundances determined with the ``NLTE'' parameters for reference.
Additional NLTE corrections to individual elements will also be discussed
were necessary and available.  Lastly, we note this analysis did
not consider possible 1D-3D effects and the possible systematic
biases introduced by our assumption of plane-parallel model
atmospheres.  3D effects can be large for stars of this metallicity,
but exploration of them except for specific elements is beyond the scope of this paper.

\subsection{Element Abundance Analysis}\label{abundances}
Abundances for the following elements were determined using measured
EWs and the LTE and ``NLTE'' sets of stellar parameters described above:
Na\,I, Mg\,I, Si\,I, Ca\,I, Ti\,I, Ti\,II, Cr\,I, Co\,I, Ni\,I and Zn\,I.  Each
line measurement was visually inspected and strong outliers were
removed.  Spectrum synthesis was used to determine abundances for Li,
C, Sc\,II, Mn\,I, and Sr\,II.  Sets of three
synthetic spectra of varying element abundance were generated using
MOOG and plotted over the observed spectrum.  The synthetic spectra
were convolved with a Gaussian to match the resolution of the data and
the continuum level was adjusted where necessary.  The element
abundance was then varied until the best match was found.  This was
done by visually evaluating the residuals of the (synthetic $-$
observed) data.

Upper limits to element abundances were calculated based on the noise
level of the spectrum.  The $\sigma$ of the noise level can be
  assessed using the Cayrel formula: 
$\sigma \approx 1.5 \times (S/N)^{-1} \times \sqrt(\rm FWHM \times
\delta x)$, where S/N is the signal-to-noise ratio, FWHM is the typical
full-width-at-half-maximum of absorption lines in that part of the
spectrum, and $\delta x$ is the spectral dispersion
\citep{cayrel_minEW}.  Typical values were $\sim$0.15 m\AA.  
The
corresponding 3$\sigma$ upper limit EW of 0.5 m\AA\ is comparable to
the mean $\Delta$EW uncertainty of 0.6 m\AA\ discussed earlier.
However, we have opted to set the upper limit EW to 1 m\AA, the
minimum accepted value for detected lines (Table~\ref{tab_ews}).
This value was used to determine upper limits to the 
abundances of Y\,II (3788 \AA), Zr\,II (4209 \AA), 
Ba\,II (4554 \AA), La\,II (4123 \AA) and Eu\,II (4129 \AA).  
The element abundances
corresponding to these upper limit EWs were found using either the
`blends' or `abfind' routine in MOOG, for lines with and without
hyperfine or isotopic splitting, respectively.  In the case of Ba, we
adopted the r-process only isotopic ratio.
Individual line LTE
abundances are given in Table~\ref{tab_ews}.

Figure~\ref{f_synth} shows the results of this upper limit analysis,
as well as spectrum synthesis of the region around the Sr\,II 4077
\AA\ feature.  For the Sr synthesis, synthetic spectra with
log$\epsilon$(Sr) $\pm$0.3 dex around the best-fit abundance are shown
by red lines.  For all the other species, synthetic spectra with the
abundances determined from the 1 m\AA\ EW upper limits are shown as
black solid lines, while the red lines represent synthetic spectra
with abundances found using EWs of 0.5 and 1.5 m\AA.  As can be seen,
1 m\AA\ is a reasonable upper limit in each region of the spectrum.

It has been noted in the literature that the Mn\,I resonance lines at
4030\AA\ indicate systematically lower Mn abundances than do
weaker non-resonance lines \citep{cayrel2004,lai2008}.  Our
investigations of these offsets in HD~122563 and HD~140283
found an offset of $+$0.30 dex (in the sense non-resonance minus
resonance), in agreement with literature studies.  Therefore, we have
corrected the abundances determined from Mn\,I 4030\AA, 4033\AA, and
4034 \AA\ by $+$0.30 dex.  The individual Mn line abundances in
Tables~\ref{tab_ews} include this offset.  \citet{bergemann_mn} have
shown that this 0.3 dex offset can be attributed to NLTE effects on
the resonance lines.

Abundance results are presented in Table~\ref{tab_lte_abund} adopting
the LTE stellar parameters, while the results of the ``NLTE'' analysis
are shown in Table~\ref{tab_nlte_abund}.  The adopted solar abundances
are those of \citet{asplund09}.

\subsection{Analysis of the Uncertainties}\label{unc}
We evaluated the uncertainties in the stellar parameters of \cd24\ in
the following way.  \teff\ and \vt\ were adjusted until slopes were
introduced into relations of Fe\,I line abundance with E.P. and reduced
EW that exceeded tolerable levels given the 1$\sigma$ dispersion in Fe\,I line
abundances.  Surface gravity was adjusted until $\rm[Fe\,II/H]$ $-$ \feh\ =
$(\sigma^{2}_{\rm Fe\,I} + \sigma^{2}_{\rm Fe\,II})^{1/2}$.  The results
are $\Delta$\teff = 60 K, $\Delta$\logg = 0.30, and $\Delta$\vt = 0.1
\kms.  These values are consistent with those found using the
empirical relations of \citet{roederer_313stars}: $\Delta$\teff = 61 K
(40 K), $\Delta$\vt = 0.05 \kms (0.15 \kms) for the subgiant (main
sequence) star relations.  Lastly, we set $\Delta$[M/H] =
$\sigma_{\rm Fe\,I}$.

We evaluated the sensitivity of these parameters to uncertainties
  in EW measures in a Monte Carlo fashion.  Starting with the EW 
measures in Table~\ref{tab_ews}, we generated Gaussian distributions of
EWs for each line with the FWHM equal to the line's $\Delta$EW.  We
then randomly selected an EW from these distributions and generated 10
sets of EW measures.  For the smallest Fe\,I lines, if the
resulting EW was smaller than 1 m\AA, it was excluded (generally no
more than 2$-$3 lines were excluded).
Starting from the original stellar parameters (before application of
the \teff\ correction),
\teff, \vt\ and \logg\ were varied to establish ionization and
excitation balance and to remove abundance trends with line strength.
The empirical \teff\ correction was then applied and final adjustments
to \logg, \vt, and [M/H] were performed.
The difference between the resulting parameters and the LTE
parameters in Table~\ref{tab_params} for each of the 10 trials is given
in Table~\ref{par_unc}.
Considering the magnitudes of the differences, they are on average 59
($\sigma$=34) K for \teff, 0.15 ($\sigma$=0.15) dex for \logg, 0.05
($\sigma$=0.03) \kms\ for \vt, and 0.04 ($\sigma$=0.03) dex for \feh.
These values along with the uncertainties in the previous paragraph
were added in quadrature to determine the total uncertainties in the
spectroscopic parameters: 84 K for \teff, 0.34 dex for \logg, 0.11
\kms\ for \vt, and 0.11 dex for [M/H].
The sensitivity of element abundances to these parameter uncertainties
were determined by varying each parameter by its uncertainty
independently.  Table~\ref{tab_unc} gives
the abundance uncertainties for each element.
 
For the non-Fe
species, we determined the uncertainty in the abundance due to EW
error by using the 10 variations of the line list 
 and determined individual line abundances using the LTE
stellar parameters.  The mean element abundances were then compared to
those in Table~\ref{tab_lte_abund}.  Typical differences were 0.02 to
0.03 dex, but were as large as 0.14 dex in the case of the single
Zn\,I line measured.
We take the maximum of either
this difference or the standard deviation of the line abundances in Table
~\ref{tab_lte_abund} as the element abundance sensitivity to EW uncertainty.
The total sensitivity of element abundances to stellar parameter and
EW uncertainties  was found by adding all the uncertainties in
quadrature, as given in the last column of Table~\ref{tab_unc}.

\section{Results and Discussion}\label{results}

\begin{figure}[!ht]
\begin{center}
   \includegraphics[clip=true,width=10cm]{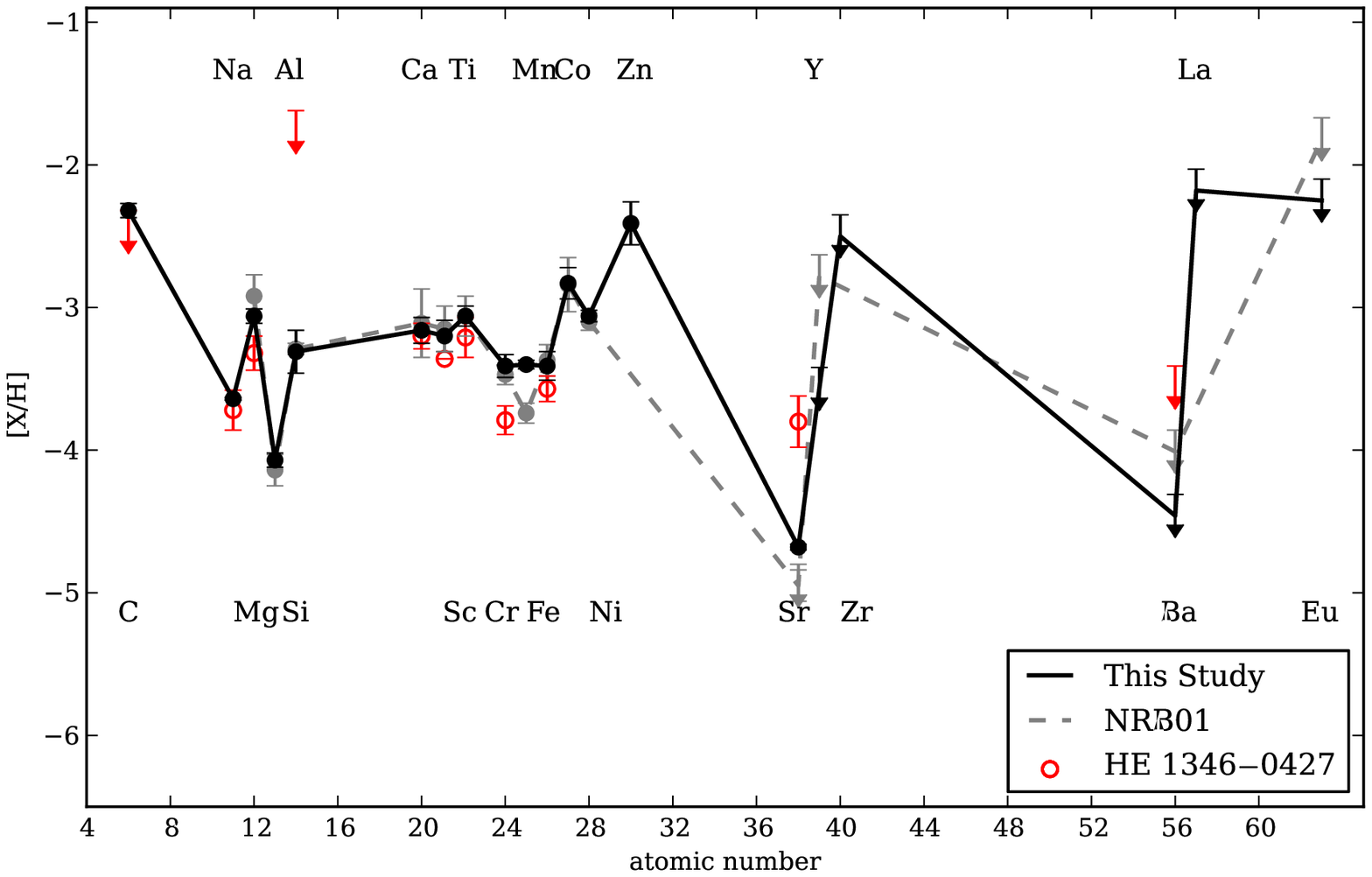} 
     \figcaption{ \label{f_nrbabundcomp}
     Comparison of element abundances for \cd24\ found in this study
     (LTE; black), compared to that of NRB01 (gray).  Upper limits are 
     indicated by
     arrows; errorbars indicate 1$\sigma$ line-by-line abundance
     dispersions.
     For comparison, we show in red the abundance pattern of
     HE~1346$-$0427 from \citet{yong13_II}, which has similar stellar
     parameters to \cd24\ and element abundances typical for stars of
     its metallicity.  See text for more information.}
 \end{center}
\end{figure}

\subsection{Comparison to literature results}\label{disclit}
In this section, we compare our abundance results to those of different
studies from the literature: NRB01,
\citet{ishigaki2010,ishigaki2012,ishigaki2013} 
and
\citet{teff_calib}.  Figure~\ref{f_nrbabundcomp} illustrates the overall
abundance ($\rm [X/H]$) pattern found here and in NRB01, with their
abundances placed on our solar abundance scale.  We also show for
comparison a star with similar stellar parameters from the sample of
\citet{yong13_II}.  This star, HE~1346$-$0427, has
\teff/\logg/\vt/\feh\ = 6255/3.69/1.40/$-$3.57, and has an abundance
pattern typical for stars of its metallicity (see Figure 43 of
\citealt{yong13_II}).  As can be seen, \cd24\ has a similar abundance
pattern to this star for elements up to the Fe-peak, apart from C and
Al, for which HE~1346$-$0427 only has upper limits.  This confirms that 
apart from C and the neutron-capture species,
\cd24\ also has element abundances typical for stars of its metallicity.

For the
elements with atomic number $Z < 29$ in common, the agreement of our
results with NRB01 is
excellent, apart from Mn.  NRB01 measured only the Mn resonance lines,
and do not include any systematic correction to the abundances.  The
$\sim$0.3 dex discrepancy between their abundance and ours can be
entirely explained by the lack of such correction.
For the neutron-capture elements, NRB01 provided upper limits for Sr,
Y, Ba and Eu.  For all but Sr, we have been able to lower the upper
limits for these species by a minimum of 0.40 dex, or a factor of 2.5.
Our detections of the 4077 \AA\ and
4215 \AA\ features yield a larger Sr abundance than found by NRB01 by
nearly 0.2 dex.  We can reproduce their upper limit using their
  stellar parameters.   

\begin{figure}[!ht]
\begin{center}
   \includegraphics[clip=true,width=8cm]{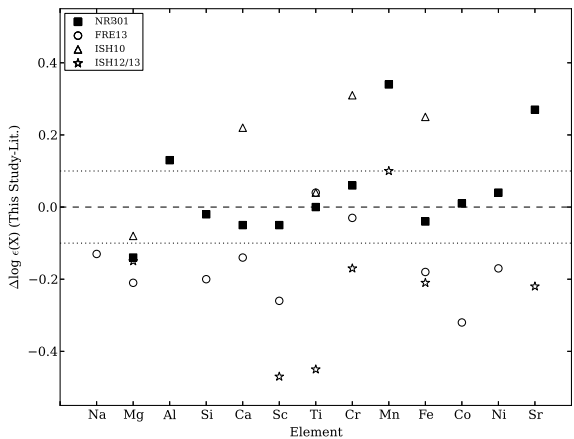} 
     \figcaption{ \label{f_litabundcomp}
     Abundance differences (in the sense This Study $-$ Literature)
     between the present work and three previous studies.  Dotted
     lines indicate $\Delta$ log $\epsilon$(X) = $\pm$0.1 to guide the
     eye.  Our results
     agree with those of NRB01 within 0.1 dex for most elements.  See
     text for more information.}
 \end{center}
\end{figure}

Figure~\ref{f_litabundcomp} illustrates the comparison to NRB01
another way, along with comparisons to 
\citet{teff_calib}, \citet{ishigaki2010} and \citet{ishigaki2012,
  ishigaki2013}
 (placed on our abundance scale, again relative to
our LTE abundances).  Here we distinguish between the results of
\citet{ishigaki2010} from the more recent papers, as the analyses (of
the same spectrum) are independent and use very different stellar parameters.
In Figure~\ref{f_litabundcomp}, the difference in log $\epsilon$(X), 
in the sense
(This Study $-$ Literature) is shown for each element.  Again,
the good agreement with NRB01 (black squares) is obvious, apart from
the elements already discussed.  

Generally, our element abundances are
lower than those found by \citet{teff_calib} and
\citet{ishigaki2012,ishigaki2013}, and higher than those
found by \citet{ishigaki2010}.  The difference with \citet{teff_calib}
can largely be explained by the smaller EW measures in this study.
The differences with \citet{ishigaki2010,ishigaki2012,ishigaki2013} are likely due to the very
different atmospheric parameters used in those works (see
Figure~\ref{f_hrd}).  Adopting the stellar parameters of
\citet{ishigaki2012,ishigaki2013} 
resulted in abundances within 0.1 dex agreement with their
values. This is similarly the case when using the \citet{ishigaki2010}
parameters, though abundance discrepencies greater than 0.2
dex remained for Mg and Ti. 
Interestingly,
\citet{ishigaki2010} 
present a Ba abundance for \cd24, based on a 1 m\AA\ EW of the
4554 \AA\ feature (they do not specify it as an upper limit).  We can
reproduce their abundance, log $\epsilon$(Ba) $\approx -2.8$, adopting this EW
and their stellar parameters. However, as we cannot see a clear Ba
absorption feature of this size in our spectrum, we prefer to determine an upper
limit as already described. 

\subsection{Discussion of individual elements}
  
To place
the abundances of \cd24\ found in this work in context of other known
main sequence and turn-off stars of comparable metallicity, we have collected
element abundances for 
stars having stellar paramters
within the ranges 5900 K $\le$ \teff\ $\le$ 6500 K, 3.6 $\le$ \logg\ $\le$
4.8, and \feh\ $\le -3.0$ from the works of
\citet{yong13_II,roederer_313stars,cohen2013,aoki2013}.  We also
include, without any selection criteria, the turn-off star
samples of \citet{bonifacio09, behara2010, bonifacio2011,
  caffau_2011gto, caffau2013} and \citet{toposI}.  In all cases, literature
abundances have been placed on the \citet{asplund09} solar abundance scale.

\subsubsection{Lithium}
\begin{figure}[!ht]
\begin{center}
   \includegraphics[clip=true,width=8cm]{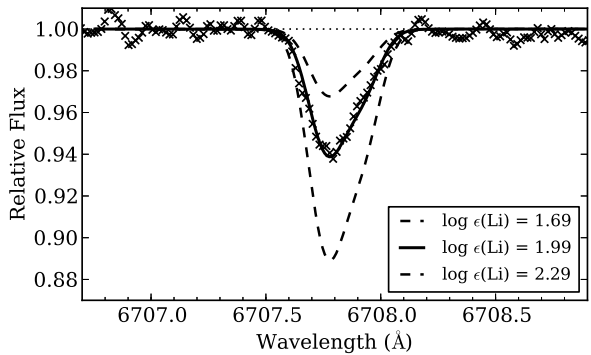} 
     \figcaption{ \label{f_lisynth}
     The Li 6707\AA\ doublet in \cd24 (crosses), 
     with the best fit LTE Li abundance
     indicated by a solid line.  Dashed lines show synthetic spectra
     with $\Delta$ log $\epsilon$(Li) = 0.3 dex around the best value.}
 \end{center}
\end{figure}

Figure~\ref{f_lisynth} illustrates spectrum synthesis of the Li~I
6707 \AA\ doublet in \cd24.  The measured EW of the feature is 18.5 m\AA.
The lithium abundance of \cd24\ has been subject to previous
study \citep{primas,aoki2009li,hosford09,melendez2004,melendez2010}.  
The LTE abundance found here, log$\epsilon$(Li) =
1.99, is in excellent agreement with that of \citet{primas}.
As the abundance of Li is sensitive to \teff, we see best agreement
with literature studies that adopted similar \teff\ values (that of
\citealt{primas} is 6300 K).  The Keck spectra used in this work
were analyzed in \citet{melendez2004} and \citet{melendez2010}.  Our measurement of the Li
feature is in excellent agreement with theirs, 18.6 m\AA, and the 0.30
dex abundance difference can be attributed to their \teff\ being $\sim$225 K
hotter than ours.

We determined the NLTE correction to this Li abundance using the grid
of \citet{lind_li} via the ``INSPECT'' website: (NLTE$-$LTE) $\Delta$
log$\epsilon$(Li) = $-$0.05.  This correction is the same regardless
of whether we use the LTE or ``NLTE'' stellar parameters, as the
\teff\ is identical in both cases.

The Li abundance patterns of unevolved extremely metal-poor stars such
as \cd24\ have been explored to investigate the behavior of the Spite plateau in the
low-metallicity regime, and our result does not add anything new to
the discussion (e.g., \citealt{spite_lithium_pl82,
  ryan_li_01, bonifacio07, melendez2004, sbordone2010, melendez2010}).
We therefore note that our Li abundance for \cd24\ is very consistent
with those of other stars of similar \teff\ and \feh\ in
\citet{sbordone2010}, and refer the reader to that paper for details
(see also, e.g., \citealt{melendez2010}).

\subsubsection{Carbon}

\begin{figure*}[!ht]
\begin{center}
   \includegraphics[clip=true,width=14cm]{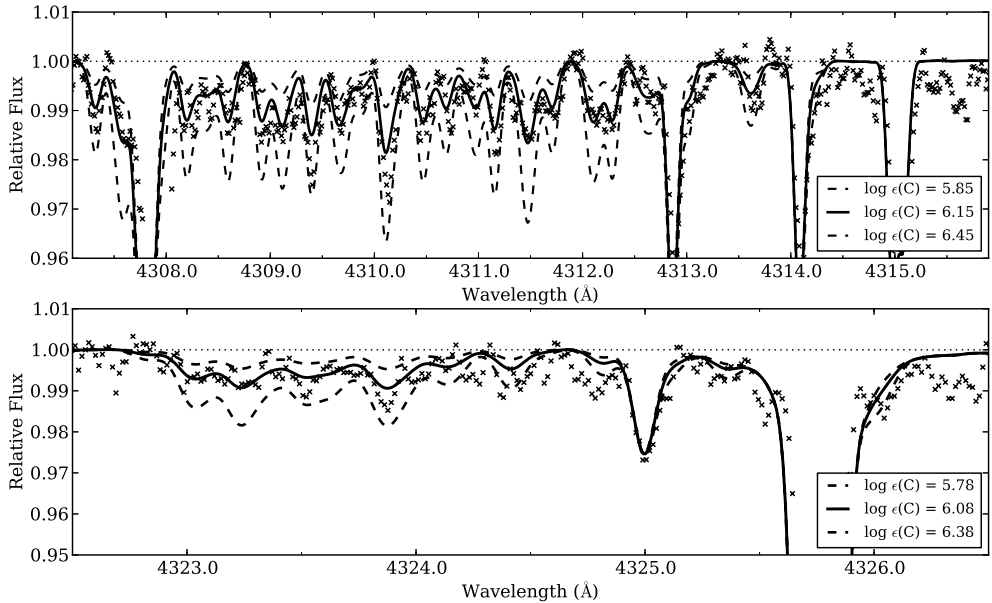} 
     \figcaption{ \label{f_csynth}
     Spectrum synthesis of the G-band CH feature in \cd24.  Carbon is
     definitely detected in this spectrum (points).  The best fit LTE
     abundance in the two regions 
     is indicated by the bold solid line.
     Abundances $\pm$0.30 dex are indicated by the dashed lines.}
 \end{center}
\end{figure*}

The CH G band is clearly detected in our spectrum of \cd24, 
as can be seen in Figure~\ref{f_csynth}.  The best fit LTE
carbon abundance based on the G band is log $\epsilon$(C) = 6.12$\pm$0.05, or
$\rm[C/Fe] = +1.10$.  Using the ``NLTE'' stellar parameters, $\rm
[C/Fe] = +0.83$.
To our knowledge, this is the first detection of
the CH G band in this star.  \citet{fabbian2009} determined a C
abundance for \cd24\ based on the EW measures of two C\,I lines in the
infrared.  They reported LTE log $\epsilon$(C) = 5.81, or $\rm[C/Fe] =
+0.59$ on the \citet{asplund09} solar abundance scale.  It is well
established in the literature that abundances determined from
molecular and atomic C features can greatly differ, due to
susceptibility to NLTE and/or 3D effects \citep{asplund_araa}.  In
particular, NLTE corrections to the C\,I features used by
\citet{fabbian2009} can be as large as $-$0.4 dex for a turn-off star of
\cd24's metallicity.  Likewise, abundances from CH features must be
decreased by as much as 0.6 dex to correct for 3D effects \citep{asplund_araa}.  To assess
whether such corrections can bring our CH abundance in better
agreement with the C\,I result from \citet{fabbian2009}, we have made
use of their EW measures to place their abundances on our scale.

First, we confirm 
their C\,I abundance using their EWs and stellar parameters
(originally from \citet{israelian2001}; see Table~\ref{tab_params})
with MOOG and a Castelli-Kurucz model atmosphere:
log $\epsilon$(C) = 5.80$\pm$0.04 (s.d.).  
C\,I abundances using our LTE and ``NLTE'' stellar parameters are
shown in Tables~\ref{tab_lte_abund} and \ref{tab_nlte_abund}.
Considering the abundances using the LTE stellar parameters, the C\,I
abundances are $\sim$0.4 dex lower than the CH carbon abundances.
According to \citet{asplund_araa}, this is expected if 3D effects
are present for CH.  However, the C\,I and CH abundances agree within
0.15 dex when our ``NLTE'' stellar parameters are used
(Table~\ref{tab_nlte_abund}).

Also shown in Tables~\ref{tab_lte_abund} and \ref{tab_nlte_abund} are
the CH and C\,I abundances corrected for 3D and NLTE effects,
respectively.  Here, we have applied a $-$0.6 dex correction to the 1D
CH abundances, following \citet{asplund_araa} (see also \citet{bonifacio09}).  For C\,I abundances,
we adopted the NLTE corrections calculated by \citet{fabbian2009} for
\cd24, assuming an {\it S$_{\rm H}$} = 1 scaling factor to collisions
with neutral hydrogen atoms (see their Table 3). With these
corrections applied, the 3D CH and NLTE C\,I abundances calculated
with the LTE stellar parameters are in good agreement: $\rm
[C/Fe] \sim +0.50$.  However, when using the ``NLTE'' stellar parameters,
there is a $\sim$0.15 dex disagreement, this time with the C\,I
abundances being larger.

It is possible that it is inappropriate to use the same 3D and NLTE
abundance corrections to the abundance results from both the LTE and
``NLTE'' stellar parameters, as the \logg\ differs by 0.33 dex, and
indicate very different evolutionary states for \cd24.  It complicates
the interpretation of the agreement/disagreement of molecular and
atomic carbon features.  According to its
LTE CH abundance ($\rm[C/Fe] = +1.1$), it qualifies as a CEMP star
according to the definition of 
\citet{ARAA} ($\rm[C/Fe] > 1$) and that set by \citet{aoki_cemp_2007}
($\rm[C/Fe] > 0.7$; the NLTE CH abundance also meets this definition).  
However, the C\,I measures of \citet{fabbian2009} and
appropriate 3D corrections indicate $\rm [C/Fe] < 1$, excluding it
from the CEMP population, though it should be stressed the CEMP
definitions used in the literature are based on 1D, LTE C abundances.

\begin{figure}[!ht]
\begin{center}
   \includegraphics[clip=true,width=8cm]{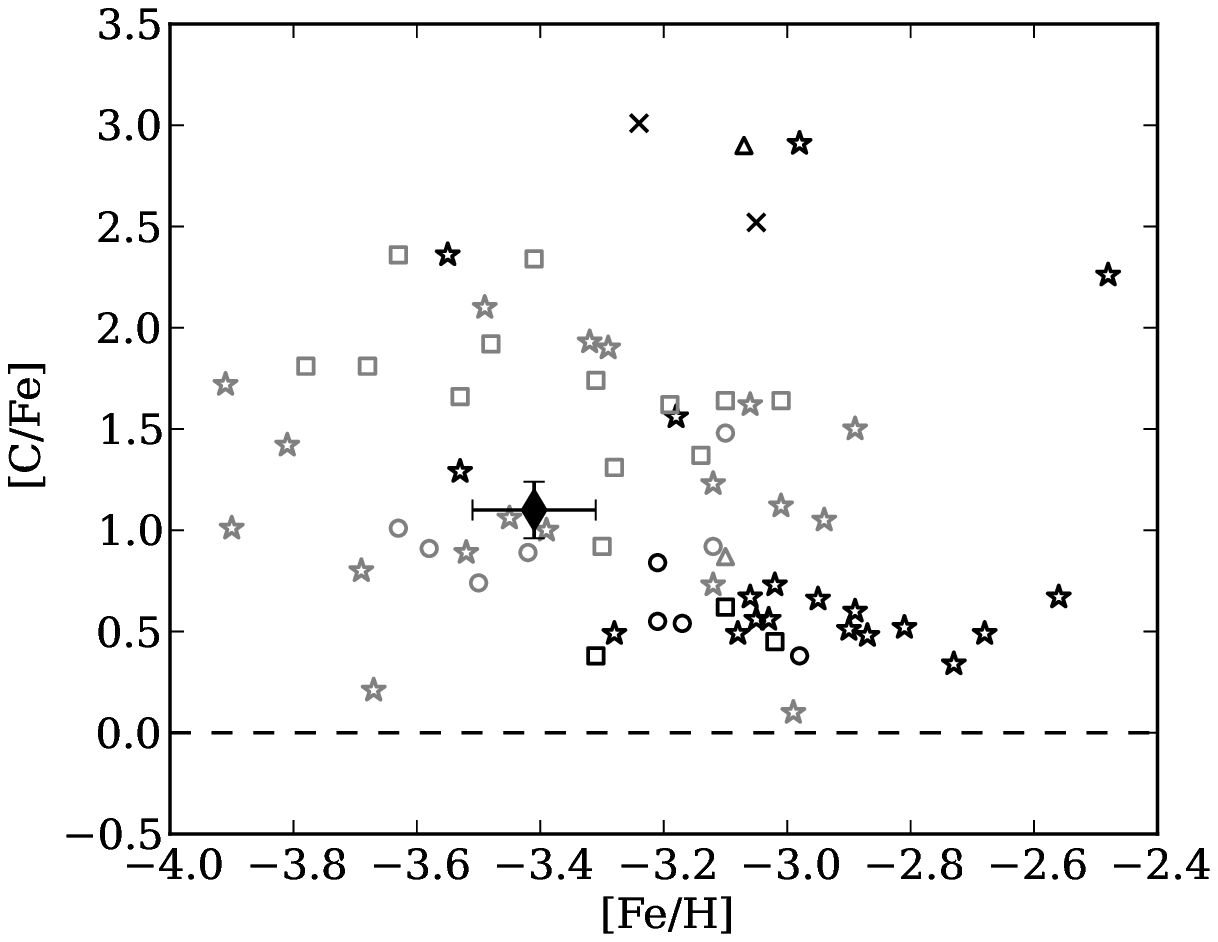} 
     \figcaption{ \label{f_cfe}
     LTE $\rm [C/Fe]$ versus \feh\ for \cd24\ (black diamond) and literature
       stars.  Open circles: \citet{roederer_313stars} (R14); open
       triangles: \citet{cohen2013} (C13); open squares:
       \citet{yong13_II} (Y13);
       crosses: \citet{aoki2013} (A13). The open stars represent results
       from \citet{bonifacio09, behara2010, caffau2013, toposI} (Bon).  In
       all cases, black symbols indicate measurements, while gray
       symbols indicate abundance upper limits.}
 \end{center}
\end{figure}

Figure~\ref{f_cfe} shows our 1D, LTE [C/Fe] ratio relative to those of
other turn-off stars from the literature.  Though most of the C
abundances for unevolved stars with \feh\ $\lesssim -3.2$ are upper
limits, there is a clear indication that the dispersion in [C/Fe]
increases with decreasing \feh.  It is also well-established that the
fraction of stars that exhibit enhanced [C/Fe] ratios increases with
decreasing \feh\ \citep{cohen,lucatello2006,carollo12,yong13_III,YSLee_cemp,placco14}.  
While \cd24\ qualifies as a
CEMP star, other stars at comparable metallicity can
have much larger enhancements.  CEMP stars can be further classified
in subcategories, depending on whether or not they exhibit
enhancements in other elements (e.g., the neutron-capture species;
CEMP-s, CEMP-r/s).  As shown in the next sections, \cd24\ has normal
[X/Fe] ratios for other species and lacks enhancements in
neutron-capture elment abundances.  Therefore, it can be classified as a
CEMP-no (``no'' for ``normal'') star \citep{ARAA}.

\citet{placco14} recently presented a comprehensive compilation of
carbon abundances for extremely metal-poor stars in the literature in
order to evaluate how the fraction of CEMP stars varies as a function
of metallicity.  Their analysis included corrections to C abundances
as a function of stellar evolutionary state to account for the
variation of C due to internal mixing as a star evolves along the
giant branch in the HR diagram.  As an unevolved star, \cd24\ does
not need such a correction, and it can be added to the sample of stars
with \feh\ $\leq -$3 and [C/Fe] $\geq$ 1.  Based on their literature sample,
\citet{placco14} found 53/168 = 32\% stars meeting this criterion.
The addition of \cd24\ to this set changes this statistic by
only a fraction of a per cent.  Considering stars with \feh\ $\leq
-$3.3, the fraction increases to 41\% (35/85) with the inclusion of \cd24.

In summary, the 1D, LTE carbon abundance as measured from the CH
  G-band in \cd24\ indicates that it is a CEMP-no star, relative to
  comparable measurements (e.g., in 1D, LTE) of other metal-poor stars in
  the literature\footnote{In comparison to G64$-$12, another
  star with very similar atmospheric parameters, \cd24\ has a
  $\sim$0.5 dex higher $\rm[C/Fe]$ ratio, as determined from a 1D, LTE
  analysis of the CH G band \citep{heresII}.}.  
  We reiterate, however, that the 3D CH abundance, as
  well as the abundances of C\,I lines and abundances determined using
  the ``NLTE'' stellar parameters for this star do not fulfill the
  CEMP star criterion; however, the \citet{ARAA} and \citet{aoki_cemp_2007}
  definitions would need to be ``translated'' to be applicable to
  abundances other than those obtained with 1D/LTE models before
  arriving at a final conclusion.

\subsubsection{Oxygen}

Although we do not detect any oxygen absorption features in our spectrum of
\cd24, multiple measures of O in \cd24\ exist in the literature, and
so we include a discussion of them for completeness.
\citet{fabbian2009} report a robust detection of a weak (1.7 m\AA)
O\,I feature at $\lambda$7772.  Abundances measured from OH features
in the near-UV have been reported by \citet{israelian2001} and
\citet{rich09}.
As for carbon, we place the O measurement by \citet{fabbian2009} on
our abundance scale by using their EW and our stellar parameters
(Tables~\ref{tab_lte_abund} and \ref{tab_nlte_abund}).
Using the LTE stellar parameters, we find log $\epsilon$(O) = 6.12
($\rm[O/Fe] = +0.84$),
which is in good agreement with \citet{fabbian2009}: 6.24.
Literature measurements of OH lines result in much larger abundances:
\citet{israelian2001} reported log $\epsilon$(O) = 6.85$\pm$0.09, while
\citet{rich09} found log $\epsilon$(O) = 6.45$\pm$0.15.  As both these
studies performed spectrum synthesis, we can not reproduce their
measures.  However, as for carbon, 3D and NLTE effects must be
considered for OH and O\,I abundance measures, respectively.  Using
the NLTE correction calculated by \citet{fabbian2009} for \cd24, the
O\,I abundance becomes $\rm [O/H]_{O\,I} = -2.99$ (LTE parameters) or
$-2.77$ (NLTE parameters).  Applying a $-$0.9 dex correction to the OH
abundances \citep{asplund_araa}, $\rm [O/H]_{OH} = -2.74$
\citep{israelian2001}, or $\rm [O/H]_{OH} = -2.84$ \citep{rich09}.
Considering the different \teff\ scales of the different studies,
these results are in good agreement.

\begin{figure*}[!ht]
\begin{center}
   \includegraphics[clip=true,width=14cm]{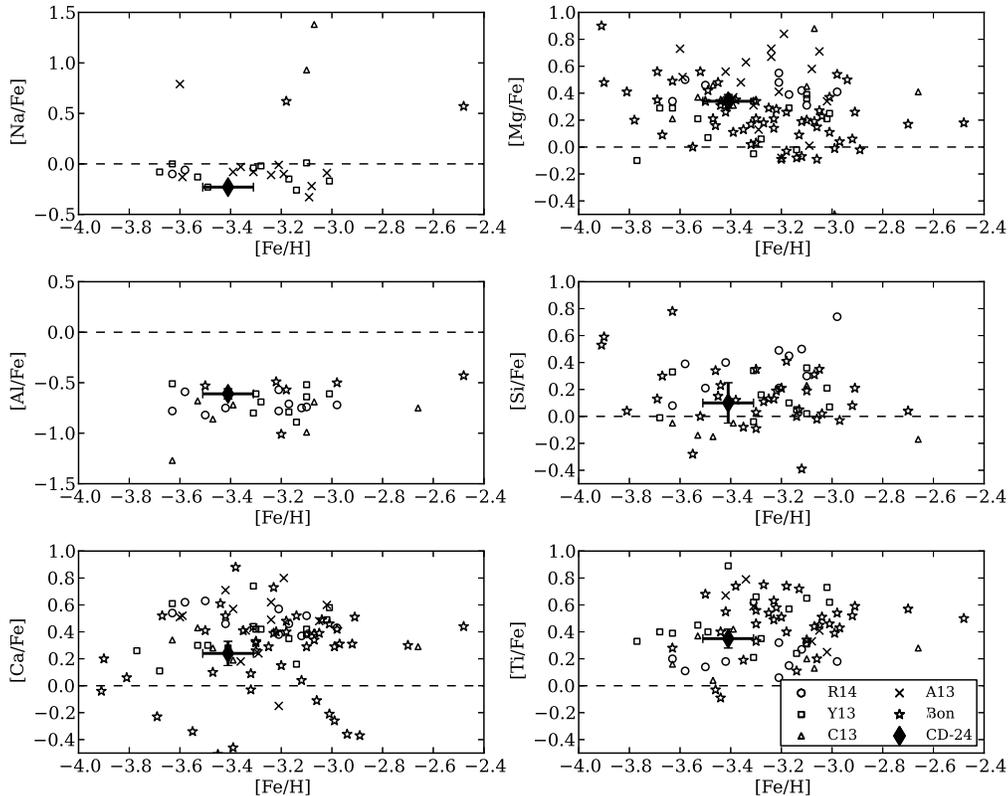} 
     \figcaption{ \label{f_alphafe}
     Light and $\alpha$-element LTE [X/Fe] ratios for \cd24\ and
     literature sample.  Symbols same as in Figure~\ref{f_cfe}.  We
     have added 0.65 dex to the Al abundances of \citet{cohen2013} to
     place their NLTE abundances on the same scale as the LTE
     abundances considered here. The Ti abundances shown are those
     determined from Ti\,II lines.}
 \end{center}
\end{figure*}

Few stars in the literature studies we are comparing to in this work
(see references in Figure~\ref{f_cfe}) provide O abundances for
turn-off stars, so we do not show plots of [O/Fe] versus [Fe/H] here.
We refer to the reader to \citet{israelian2001} and
\citet{fabbian2009} to see the oxygen abundance of \cd24\ in the
context of other extremely metal-poor turn-off stars.  Briefly, its
[O/Fe] (as measured by both O\,I and OH species) is in good agreement
with general trends shown by other halo stars.

\subsubsection{Light and $\alpha$-elements}

Figure~\ref{f_alphafe} shows LTE [X/Fe] versus \feh\ for the light
elements Na and Al, as well as the $\alpha$-elements for \cd24\ and
literature
stars\footnote{\citet{bonifacio2011,caffau_2011gto,caffau2013,toposI} present
abundances of both Ca\,I and Ca\,II for their stellar samples.  Here
we consider only their Ca\,I abundances, to be consistent with this
and other literature studies considered.}.  
As can be seen, \cd24\ exhibits [Na/Fe] and [Al/Fe]
ratios similar to those of other stars at similar metallicity.  
We determined NLTE corrections for the two Na\,I lines considered here
using the grid of \citet{lind_na} in the ``INSPECT'' website.  They
are (in the sense NLTE$-$LTE) $-$0.07 and $-$0.06 dex for the
5889\AA\ and 5895\AA\ lines, respectively.  
\citet{al_nlte} found NLTE corrections
for Al abundances determined from the 3961\AA\ line to be as large as
$\sim$0.65 dex for stars of similar evolutionary state to \cd24.
While their stellar sample did not contain any stars with \feh $< -$3,
corrections for stars like \cd24\ are likely to be at least of the same
magnitude.  Such corrections would shift the [Al/Fe] ratios shown in
Figure~\ref{f_alphafe} to roughly the solar ratio.  Solar ratios of
[Al/Fe] at low \feh\ are much more consistent with predictions of
chemical evolution models, as noted by \citet{al_nlte} and others. 

\cd24\
also has typical enhanced [$\alpha$/Fe] ratios ($< \rm [\alpha/Fe] >$
= 0.35) for a halo star.  For the stars shown in
Figure~\ref{f_alphafe},
abundance enhancements are typically largest
for Mg and Ti, and are less pronounced for Si (though a few stars have
[Si/Fe] $> +0.5$).  While many stars have [Ca/Fe] $> +0.5$, some
exhibit sub-solar [Ca/Fe] ratios.  Indeed, the number of extremely
metal-poor stars that do not show enhanced [$\alpha$/Fe] ratios has
grown in recent years (e.g., \citealt{caffau2013, toposI}), and their
existence indicates a certain degree of inhomogeneity in the chemical
enrichment of the early Galaxy, or the sub-halos that built up the
Galaxy.

\begin{figure*}[!ht]
\begin{center}
   \includegraphics[clip=true,width=15cm]{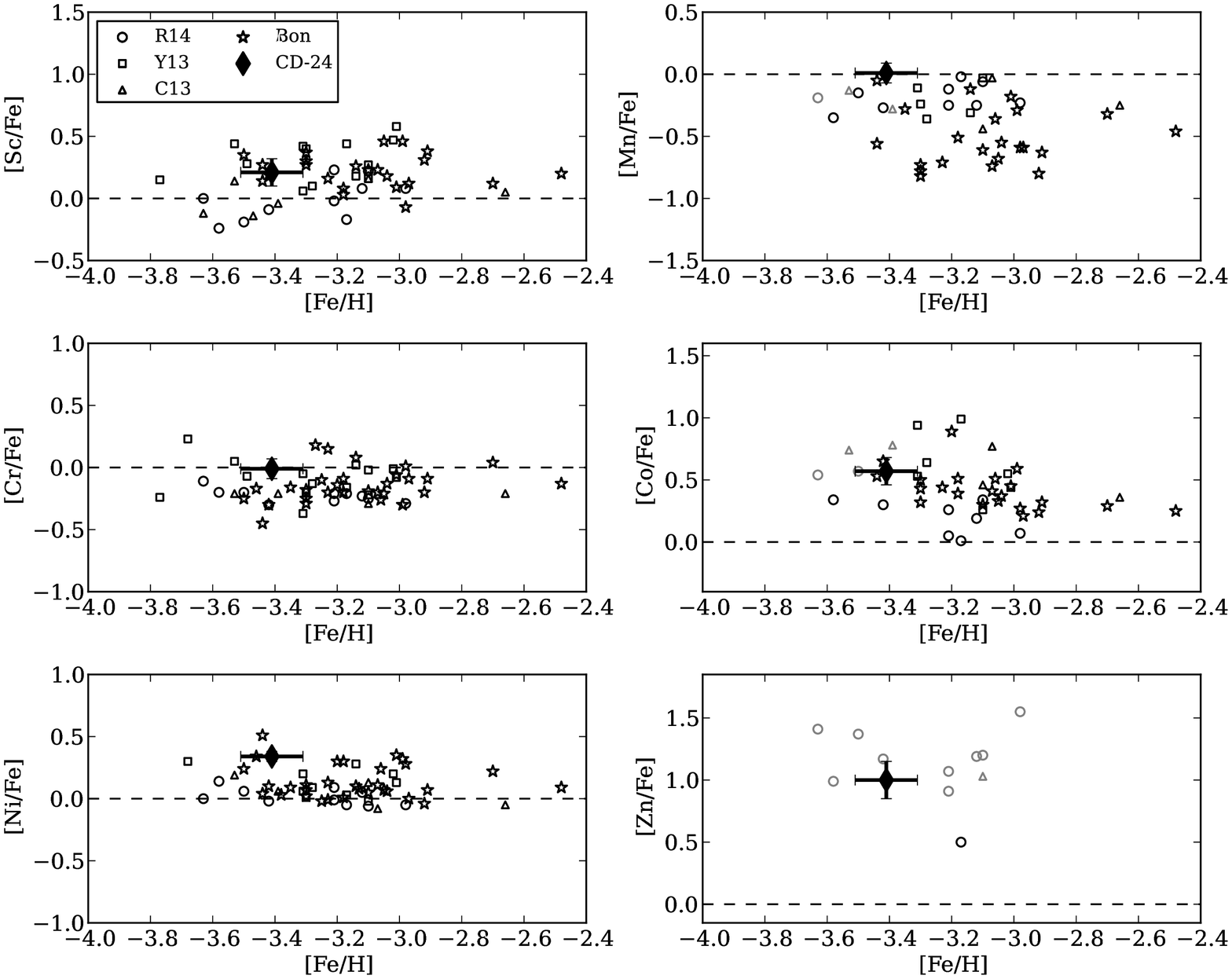} 
     \figcaption{ \label{f_fepeak}
     Same as Figure~\ref{f_alphafe} but for the Fe-peak elements.}
 \end{center}
\end{figure*}

Metal-poor star $\alpha$-element abundance determinations are also
susceptible to NLTE effects.  For Mg and Ti, the NLTE corrections are
relatively small, $\sim$ $+$0.1 and $-$0.05\footnote{This correction
  is found for Ti\,II lines, which are more numerous and reliably
  measured than Ti\,I lines in metal-poor star spectra.  Consequently,
Ti\,II is more frequently used to determine a star's Ti abundance.}
dex, respectively \citep{gehren2004_nlte,bergemann_ti}.
\citet{mashonkina_ca} found NLTE corrections for Ca\,I lines that
varied from $+$0.10 to $+$0.29 dex for a warm metal-poor star such as
\cd24.  Corrections for Si\,I abundances are even larger: for
G64$-$12, a star with similar parameters to \cd24,
\citet{shi_si} found the NLTE correction for the 3905\AA\ line
abundance to be $+$0.25 dex.  The general impact of these corrections
would be to increase the level of $\alpha$-element enhancement of the
stars in Figure~\ref{f_alphafe}, but would not necessarily change the
interpretation of the data.  The [Ca/Fe]-poor stars, for example,
would still remain [Ca/Fe]-poor relative to the general halo star population.

\subsubsection{Fe-peak elements}

LTE [X/Fe] ratios versus \feh\ are shown in Figure~\ref{f_fepeak} for
scandium and the Fe-peak elements.
Again, the abundance pattern of \cd24\ is similar to that of stars of
comparable metallicity.  The relatively large [Mn/Fe] ratio for
\cd24\ compared to that of the literature sample can be entirely
explained by the 0.3 dex offset applied to the resonance line
abundances; such a correction was not performed on the literature
results.

The [Ni/Fe] ratio of \cd24, $+$0.34 dex (LTE), is higher than the
typical halo star value that is approximately solar.  However, as can
be seen in Figure~\ref{f_fepeak}, several stars with \feh\ $< -$3 show
enhanced [Ni/Fe] ratios.  Zn abundances are difficult to determine in
metal-poor star spectra, as only a couple weak Zn\,I lines are
present.  While these lines can be accurately measured in our spectrum
of \cd24, the Zn abundances for most of the metal-poor turn-off stars
in the literature are upper limits.

NLTE corrections have been determined for Mn, Cr and Co.
\citet{bergemann_mn} found corrections of $+$0.3 dex are required for
the Mn resonance lines, and that is accounted for by our
empirically-determined correction.  Corrections of order $+$0.35-0.40
dex for Cr\,I lines were found for G64$-$12 by \citet{bergemann_cr}.
Given that \cd24\ has similar stellar parameters to G64$-$12, these
corrections are applicable in this case.

\begin{figure*}[!ht]
\begin{center}
   \includegraphics[clip=true,width=14cm]{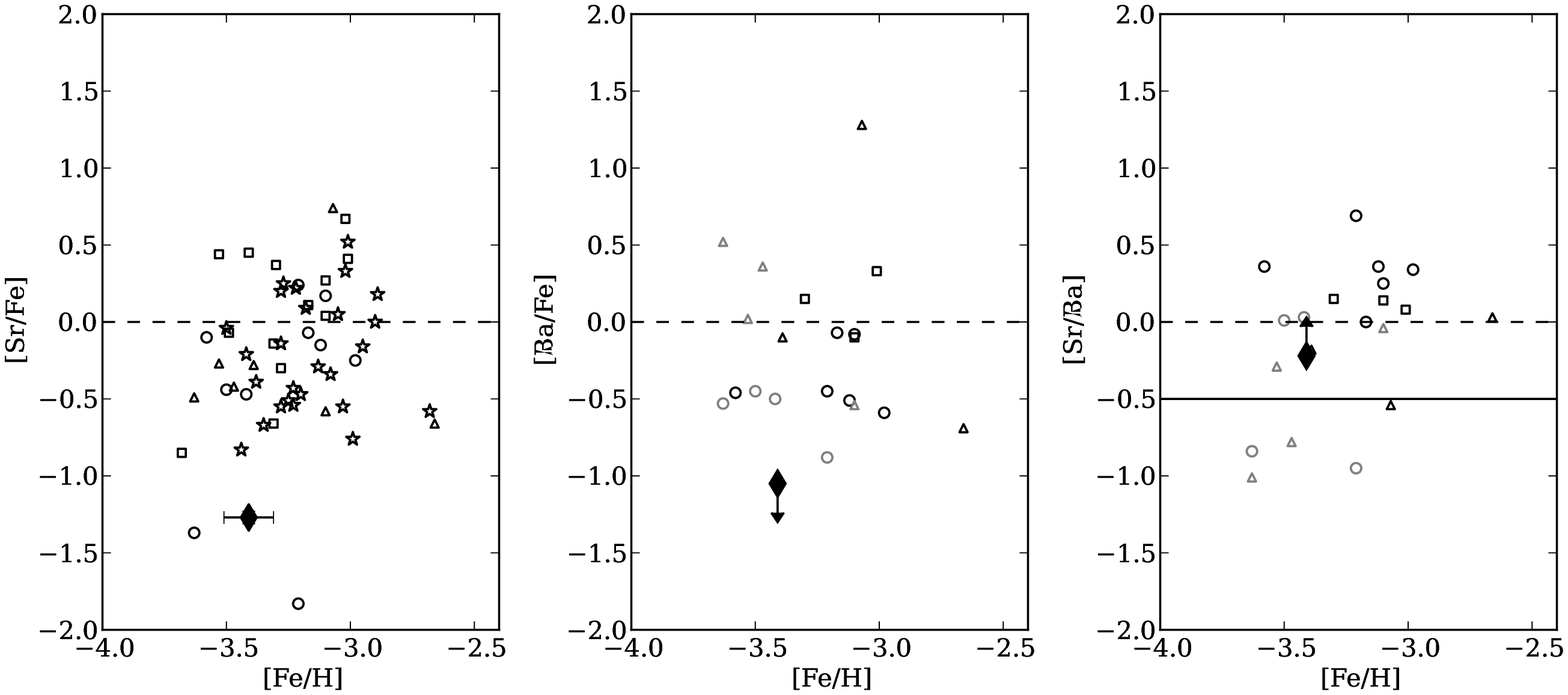} 
     \figcaption{ \label{f_ncap}
     Same as Figure~\ref{f_alphafe} but for the neutron-capture
     elements Sr and Ba.  The right panel shows the ratio
     of light-to-heavy neutron-capture element abundances, [Sr/Ba],
     versus [Fe/H]. The pure r-process [Sr/Ba] ratio is indicated by
     the solid line \citep{simmerer2004}.}
 \end{center}
\end{figure*}

Corrections for Co\,I lines can be as large as $+$1 dex for cool,
evolved metal-poor stars, but in the case of \cd24, the corrections are
$\sim$0.65 dex \citep{bergemann_co}.  Such corrections would place the
[Co/Fe] ratios shown in Figure~\ref{f_fepeak} close to [Co/Fe]$\sim$1
and above.  As noted by \citet{bergemann_co}, such ratios are at odds
with chemical evolution models which use metallicity-dependent
supernova yields (e.g., \citealt{kobayashi2006}).

\subsubsection{Neutron-capture elements}

Of the neutron-capture species that can generally be measured in
metal-poor star spectra, only Sr can be detected in even our high S/N
spectrum of \cd24.  As mentioned in Section~\ref{disclit}, our {\it
  measured} Sr
abundance is at slight odds with the upper limit determined by NRB01,
being larger by $\sim$0.2 dex.  This difference can be
attributed to choice of stellar parameters.  Our EWs agree well with
theirs: they adopted an upper limit EW of 4 m\AA\ for both 4077 and
4215 \AA\ Sr\,II lines, while our bona fide measures are 5.5 and 3.1
m\AA, respectively.  As can be seen in the left panel of
Figure~\ref{f_ncap}, \cd24\ exhibits one of the lowest Sr abundances
of unevolved metal-poor stars.  Although we are considering a
relatively narrow range of stellar parameter space (only dwarfs or
stars near the MSTO), a $>$1 dex
dispersion in Sr abundances can be observed.  This behavior for the
neutron-capture elements, in contrast to the $\sim$0.1 dex dispersion
seen for other element groups in the Periodic Table, has been well
remarked on in the literature \citep{2000burris,heresII,sneden_araa,psss}.

\begin{figure*}[!ht]
\begin{center}
   \includegraphics[clip=true,width=14cm]{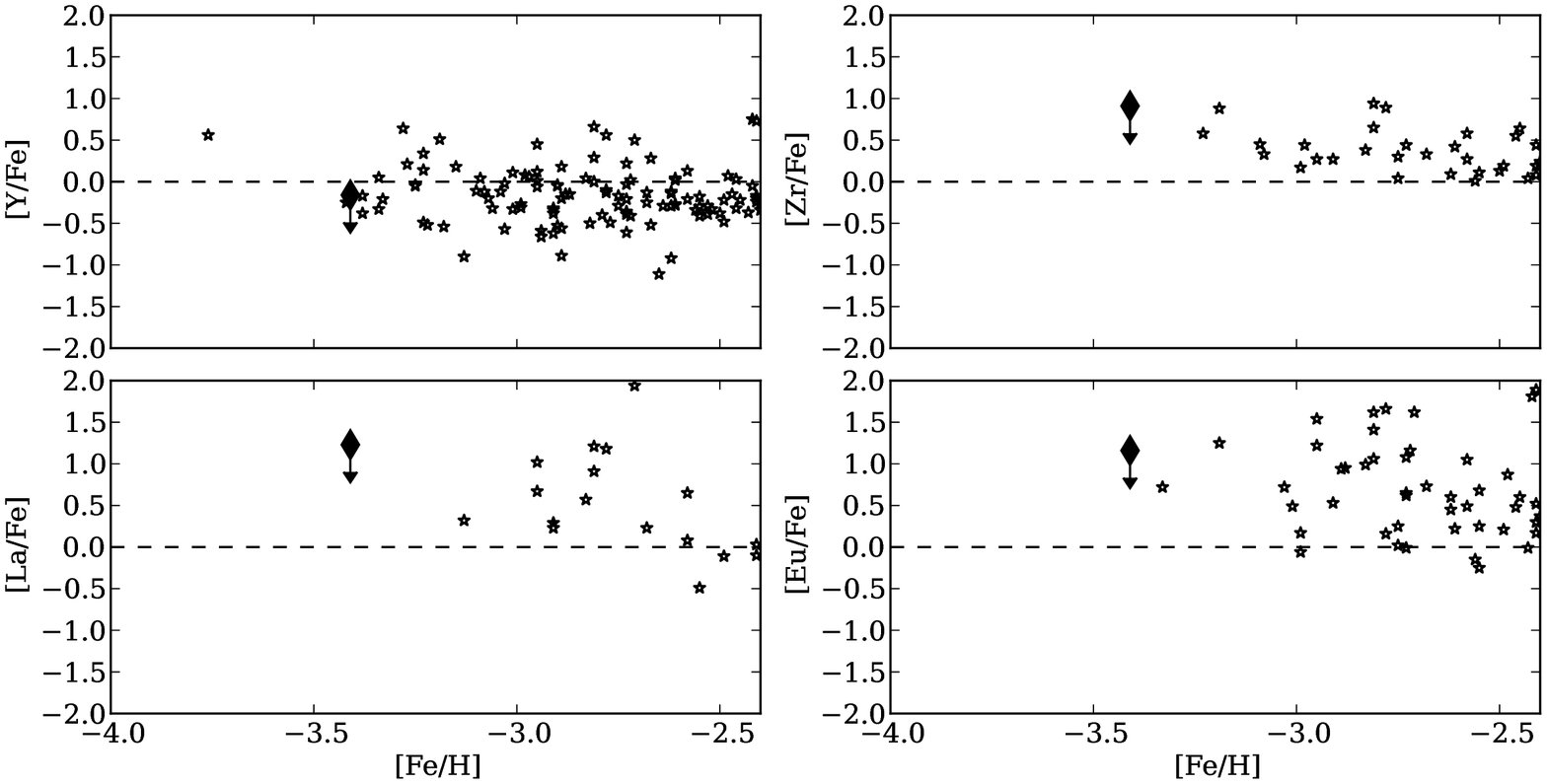} 
     \figcaption{ \label{f_otherncaps}
    The LTE upper limit [X/Fe] ratios for the elements Y, Zr, La and
    Eu in  \cd24 (filled diamonds), this time
     in comparison to the sample of \citet{heresII} (stars).  }
 \end{center}
\end{figure*}

As described in Section~\ref{abundances}, upper limits to
EWs were determined for neutron-capture species lacking
detectable absorption lines in our spectrum.  Corresponding abundance
upper limits were determined based on these EW upper limits.
Our abundance upper limit for barium, [Ba/Fe] $< -$1.05,
is a factor of 2.8 (4.5 dex) lower than the upper limit of NRB01,
and is the lowest shown in Figure~\ref{f_ncap}\footnote{We note that
  other stars not considered here can exhibit comparably low [Ba/Fe]
  abundances.  See, e.g., Figure 7 in \citet{segue1} for a recent
  comprehensive literature compilation.}.  The last panel of
Figure~\ref{f_ncap} shows [Sr/Ba] versus \feh, which indicates the
relative abundances of light to heavy neutron-capture species.  
Though \cd24\ exhibits
some of the lowest Sr and Ba abundances, the (lower limit) of their
ratio is close to solar and well above the ratio produced by the pure
r-process.  Therefore another source besides the main r-process
contributed to the Sr abundance in this star, along with that of 
most of the literature stars also considered here.  An analysis of the
dispersions in Sr and Ba abundances within a larger, homogeneous
sample of unevolved extremely metal-poor stars is the subject of a future paper.

Zr\,II, Y\,II  and La\,II abundances are not commonly included in high
resolution spectroscopic studies of metal-poor stars, except in cases
where stars show enhancement in neutron-capture element abundances or
where the data quality is exceptionally high: their spectroscopic
features are generally too weak to measure.  They are undetectable
even in our high quality data.
We include them here
because NRB01 determined an upper limit to the Y\,II abundance of
\cd24.  For Y, our spectrum allows for a $\sim$0.8 dex reduction of
the lower limit to [Y/Fe] $< -$0.16.  Our upper limits to Zr and La 
abundances in \cd24\ are [Zr/Fe] $< +$0.91 and [La/Fe] $< +$1.23,
respectively.
Our upper limit measure for [Eu/Fe], $< +$1.16, is 0.4 dex lower than
that of NRB01 ($< +$1.55 ). 

Figure~\ref{f_otherncaps} shows [X/Fe] ratios Y, Zr, La and Eu in \cd24,
this time in comparison to the large homogeneous literature sample of
\citet{heresII}, as most of the literature stars in previous figures
lack abundance measurements for these species.  We also note that the
\citet{heresII} sample is a mixture of evolved and unevolved stars, in
contrast to previous figures.  As can be seen, the upper limits
measures for \cd24\ are consistent with the Y, Zr and La abundances of
the literature data, which show varying dispersions of $\lesssim$1 dex
for Zr to $>$2 dex for La and Eu.  It is important to repeat the
caution of \citet{heresII}, however, that given the difficulty of
measuring the abundances for these species in most extremely
metal-poor stars, the stellar samples in Figure~\ref{f_otherncaps} are
biased and incomplete.  Therefore, it is difficult to draw broad
conclusions about the intrinsic abundance patterns and dispersions for these
elements in the context chemical evolution.   

More firm interpretations can be made from the Sr and Ba abundances of
extremely metal-poor stars, however, as those samples are much more
complete and unbiased.  With [Sr/H] $\approx -$4.7 and 
[Ba/H] $\leq -$4.5, \cd24\ is among the Milky Way halo stars with
the lowest neutron-capture element enhancements (e.g., Fig.\ 1 of
\citealt{roederer_ncap} and Fig.\ 7 of \citealt{segue1}).  Such low
levels of neuton-capture abundances are common in stars in ultra-faint
dwarf galaxies (e.g., \citealt{segue1} and references therein).  These
low-mass systems are thought to be the remaining analogs of the
``proto-galaxies'' that were the early building blocks of large
galaxies such as the Milky Way, as only one or two stellar generations
contributed to their chemical enrichment before star formation was
quenched.  Therefore, if low levels of neutron-capture enrichment is a
signature of these primitive systems, then the Milky Way halo stars
with low Sr and Ba abundances such as \cd24\ may have originated in such systems
\citep{fnaraa}.  Given the high proper motion of \cd24, its
kinematics may also provide clues to its origin.

Such statements are generally made regarding LTE neutron-capture
element abundances, however.  \citet{bergemann_sr} found (NLTE$-$LTE)
abundance corrections for the Sr\,II 4077 feature of $-$0.05 to 0 dex
for unevolved stars at \feh\ = $-$3, but this correction can reach
$+$0.15 at lower metallicities.  \citet{andrievsky_sr} found NLTE
corrections depended upon a star's Sr abundance as well as its stellar
parameters.  For a star with \cd24's \teff, \feh, and Sr abundance,
they found a correction of $\sim +$0.25 dex (their Figure 7).
Similarly, \citet{andrievsky_ba} found NLTE corrections to Ba
abundances measured from the Ba\,II 4554 line also depend on effective
temperature and Ba abundance (see also \citealt{mashonkina_ba_nlte}).  
They did not provide corrections for
[Ba/H] $< -$3 in their work, but for turnoff stars with similar
parameters to \cd24\ they calculated NLTE corrections ranging from
$+$0.07 to $+$0.37 dex.  
For the remaining neutron-capture element
most commonly studied in metal-poor stars, Eu, NLTE corrections are
$\sim +$0.05$-$0.10 dex for unevolved stars \citep{mashonkina2003}. 

\subsection{Effects of Internal Diffusion Processes}
 The surface abundance of a star can differ from its primordial
chemical composition due to internal diffusion processes throughout
its lifetime, such as convection, gravitational settling, and
radiative acceleration (e.g.,
\citealt{richard2002_I,korn2007,korn2009}). 
\citet{richard2002} pointed to \cd24\ specifically as an extremely
metal-poor turnoff star whose observed abundance pattern may differ 
greatly from its initial composition.  Indeed, the $\sim$500 K range
in \teff\ found for \cd24\ by different studies (Figure~\ref{f_hrd},
Table~\ref{tab_params}) indicate very different evolutionary states for \cd24,
so it is difficult to quantify the magnitude of its abundance
variations.  However, looking at Figures 10 and 11 of
\citet{richard2002}, one can get a sense of how large the abundance
variations might be for specific elements, spanning the range of
\teff\ values measured for \cd24.

For example, depending on its evolutionary state, the measured Fe
abundance of \cd24 may be as much as $\sim$0.3 dex underabundant to
$\sim$0.9 dex overabundant relative to its intitial [Fe/H].  Ca may
also be under- or overabundant by as much as $\sim$0.5 dex.  Carbon
and oxygen can be depleted by nearly 2 dex, while species such as Na,
Mg, Cr and Mn can be 0.2 to 0.4 dex below their initial values.
Alternatively, Al, Si and Ni can be 0.3-0.4 dex enhanced.  

Based on our abundance analysis (Tables~\ref{tab_lte_abund},
\ref{tab_nlte_abund}), the [X/Fe] ratios of Na, Cr, Mn and Ni are
consistent with the model predictions of \citet{richard2002}, but the
[Al/Fe] $<$ 0 and [Si/Fe]$\sim$0 ratios disagree.  In order to better
compare the abundance pattern of \cd24\ to models that include
diffusion effects, it is necessary to more tightly constrain its
evolutionary status by decreasing the dispersion in stellar parameters
found by different studies using different methods.  For now, the
discussion in this work and others of how the abundance pattern of
\cd24\ compares to other halo stars (e.g, Figures~\ref{f_cfe} to
\ref{f_otherncaps}) should be viewed cautiously until the effects of
internal diffusion are better understood.

\section{Summary}\label{conc}

We have presented a comprehensive element abundance analysis of the
canonical metal-poor turn-off star \cd24\ based on high-resolution,
high S/N archival spectra.  Though comparable in resolution to other
studies of \cd24, the extremely high S/N ($\sim$400 in some places) of
our data allow for more robust equivalent width measurements and a
factor of $>$3 improvement in the lower limits to some element
abundances.  Our analysis began with a classical spectroscopic
determination of stellar parameters with the addition of an empirical
correction to effective temperature.  We also performed an ``NLTE''
parameter determination following the method of \citet{ruchti2013}.
LTE and ``NLTE'' abundances were then determined for 20 other species
using both equivalent widths and spectrum synthesis techniques.  Our
resulting stellar parameters, metallicity, and element abundances
agree well with those of other studies, including that of NRB01.  In
particular, we have found the following:

\begin{itemize}
\item \cd24\ has \feh $= -$3.41 and \feh $= -$3.29, assuming LTE and
  NLTE, respectively.  This value is in good agreement with studies in
  the literature.  Its [X/Fe] ratios for the light, $\alpha$- and
  Fe-peak elements are comparable to those of other halo stars at
  similar metallicity.
\item Based on a clear detection of the CH G-band feature in our
  spectrum of \cd24, we have measured it to have [C/Fe] $= +$1.10.
  As it lacks enhancements in any neutron-capture element abundances,
  it can therefore be classified as a CEMP-no star according to
    the definition of \citet{ARAA}.  It is among the
  $\sim$40\% of stars with \feh $\leq -$3.3 that have [C/Fe] $\geq$1 
  \citep{placco14}.  However, the C abundance as measured by infrared
  C\,I lines \citep{fabbian2009} is $\sim$0.3 dex lower, serving as a reminder
  that 3D and NLTE effects should not be disregarded and may change
  our current understanding of the behavior of [C/Fe] versus [Fe/H]
  for extremely metal-poor stars, which generally are based on 1D, LTE
  measures of CH features in stellar spectra.
\item We have compared literature oxygen abundances for \cd24\ as
  measured by near-ultraviolet OH features and the O\,I triplet.
  Making appropriate 3D and NLTE corrections brings these O abundances
  into good agreement, and indicates that \cd24\ has $\rm [O/Fe] \sim +0.5$.
\item We have decreased the upper limit abundance estimates for
  elements Ba, Y and Eu by 0.4$-$0.8 dex compared to NRB01, and also
  provide an upper limit measure for Zr and La.  The upper limits for
  Y, Zr, La and Eu are comparable to values found in other metal-poor
  stars in the literature, though data remains sparse for some of
  these elements.
\item \cd24\ has (LTE) [Sr/H]$\sim -$4.7 and [Ba/H]$\lesssim -$4.5, which
  are among the lowest Sr and Ba abundances exhibited by extremely
  metal-poor stars in the Milky Way halo.  Such low neutron-capture
  element abundances are more characteristic of stars in ultra-faint
  dwarf galaxies, and may therefore indicate that \cd24\ originated in
  a high-redshift analog of such a system.
\item We briefly review the abundance pattern of \cd24\ relative
  to the predicted abundance variations caused by internal
  diffusion processes in a metal-poor turn-off star.  The [X/Fe]
  ratios of some elements are consistent with model predictions, but
  others disagree.  The location of \cd24\ on the Hertzsprung-Russell
  diagram must be more tightly constrained before the magnitude of
  possible abundance variations can be determined.
\end{itemize}

\acknowledgements{
The authors warmly thank A. R. Casey for the development of SMH and
for making available scripts used to combine the archival data, as 
well as for discussions about and his assistance with checking the quality of
the combined spectrum.  We are grateful to the referee for comments
that improved the presentation of these results. 
This research has made use of data obtained from the ESO Science
Archive Facility and the Keck Observatory Archive.
We also acknowledge
  use of the SIMBAD database, operated at CDS, Strasbourg, France and
  of NASA's Astrophysics Data System Bibliographic Services.  This
  work also includes data obtained from the INSPECT database (v1.0). 
  A.F.\ acknowledges support from NSF grant AST-1255160.}

\textit{Facilities:} \facility{Keck:I (HIRES)}, \facility{VLT:Kueyen (UVES)}

%===========================================================================

\clearpage

\LongTables

\pagestyle{empty}

\begin{landscape}
\begin{deluxetable*}{lccccccccc}
\tablecaption{Details of Archival Spectra of \cd24 Used in this Work\label{tab_spec}}
\tablewidth{0pt}
\tabletypesize{\scriptsize}
\tablehead{
\colhead{Filename} & 
\colhead{Instrument} &
\colhead{$\lambda$ (\AA)} &
\colhead{$R \equiv \lambda/\Delta\lambda$} &
\colhead{S/N\tablenotemark{a}$@$\,4500~\AA} &
\colhead{S/N\tablenotemark{a}$@$\,6000~\AA} &
\colhead{Exp Time (s)} &
\colhead{P.I.} &
\colhead{Prop ID / Ref.\tablenotemark{b}} & 
\colhead{UT Date} }
\startdata
ADP.2013-09-25T06:40:48.133.fits & UVES & 4780$-$6800 & 51,700 &
\nodata & 24 & 101 & Primas & 68.D-0094(A) & 2001-11-26\\
ADP.2013-09-25T06:40:48.143.fits & UVES & 4780$-$6800 & 51,700 &
\nodata & 100 & 1000 & Primas & 68.D-0094(A) & 2001-11-26\\
ADP.2013-09-25T06:40:48.237.fits & UVES & 4780$-$6800 & 51,700 &
\nodata & 118 & 1000 & Primas & 68.D-0094(A) & 2001-11-26\\
ADP.2013-09-25T06:40:48.420.fits & UVES & 4780$-$6800 & 51,700 &
\nodata & 128 & 1000 & Primas & 68.D-0094(A) & 2001-11-26\\
ADP.2013-09-25T06:40:48.520.fits & UVES & 4780$-$6800 & 51,700 &
\nodata & 196 & 1000 & Primas & 68.D-0094(A) & 2001-11-26\\
ADP.2013-09-25T06:46:55.763.fits & UVES & 4780$-$6800 & 51,700 &
\nodata & 146 & 1000 & Primas & 68.D-0094(A) & 2001-11-27\\
ADP.2013-09-25T06:46:55.890.fits & UVES & 4780$-$6800 & 51,700 &
\nodata & 156  & 1000 & Primas & 68.D-0094(A) & 2001-11-27\\
ADP.2013-09-25T06:46:55.930.fits & UVES & 4780$-$6800 & 51,700 &
\nodata & 138 & 1000 & Primas & 68.D-0094(A) & 2001-11-27\\
ADP.2013-09-26T07:38:05.573.fits & UVES & 3750$-$4970 & 53,000 & 75 & \nodata & 975 & Akerman &
073.D-0024(A) & 2004-08-07\\
ADP.2013-09-26T07:38:05.580.fits & UVES & 3750$-$4970 & 53,000 & 87 & \nodata & 975 & Akerman  &
073.D-0024(A) & 2004-08-07\\
ADP.2013-09-26T07:38:05.607.fits & UVES & 3750$-$4970 & 53,000 & 59 & \nodata & 975 & Akerman  &
073.D-0024(A) & 2004-08-07\\
ADP.2013-09-26T07:38:05.620.fits & UVES & 3750$-$4970 & 53,000 & 96 & \nodata & 975 & Akerman  &
073.D-0024(A) & 2004-08-08\\
ADP.2013-09-26T07:38:05.627.fits & UVES & 3750$-$4970 & 53,000 & 69 & \nodata & 975 & Akerman  &
073.D-0024(A) & 2004-08-07\\
ADP.2013-09-26T07:38:05.653.fits & UVES & 3750$-$4970 & 53,000 & 69 & \nodata & 975 & Akerman  &
073.D-0024(A) & 2004-08-07\\
ADP.2013-09-26T07:38:05.693.fits & UVES & 3750$-$4970 & 53,000 & 115 & \nodata & 975 & Akerman  &
073.D-0024(A) & 2004-08-07\\
ADP.2013-09-26T07:38:05.753.fits & UVES & 3750$-$4970 & 53,000 & 74 & \nodata & 975 & Akerman  &
073.D-0024(A) & 2004-08-07\\
ADP.2013-09-26T07:40:12.363.fits & UVES & 3750$-$4970 & 53,000 & 87 & \nodata & 975 & Akerman  &
073.D-0024(A) & 2004-08-10\\
ADP.2013-09-26T07:40:12.450.fits & UVES & 3750$-$4970 & 53,000 & 61 & \nodata & 975 & Akerman  &
073.D-0024(A) & 2004-08-10\\
ADP.2013-09-26T07:40:12.470.fits & UVES & 3750$-$4970 & 53,000 & 94 & \nodata & 975 & Akerman  &
073.D-0024(A) & 2004-08-10\\
ADP.2013-09-26T07:40:12.477.fits & UVES & 3750$-$4970 & 53,000 & 72 & \nodata & 975 & Akerman  &
073.D-0024(A) & 2004-08-10\\
ADP.2013-09-26T07:40:12.510.fits & UVES & 3750$-$4970 & 53,000 & 68 & \nodata & 975 & Akerman  &
073.D-0024(A) & 2004-08-10\\
ADP.2013-09-26T07:40:12.523.fits & UVES & 3750$-$4970 & 53,000 & 65 & \nodata & 975 & Akerman  &
073.D-0024(A) & 2004-08-10\\
ADP.2013-09-26T07:40:12.563.fits & UVES & 3750$-$4970 & 53,000 & 78 & \nodata & 975 & Akerman  &
073.D-0024(A) & 2004-08-10\\
ADP.2013-09-26T07:55:28.080.fits & UVES & 3750$-$4970 & 53,000 & 61 & \nodata & 975 & Akerman  &
073.D-0024(A) & 2004-09-01\\
ADP.2013-09-26T07:55:28.217.fits & UVES & 3750$-$4970 & 53,000 & 65 & \nodata & 975 & Akerman  &
073.D-0024(A) & 2004-09-01\\
ADP.2013-09-26T07:55:28.230.fits & UVES & 3750$-$4970 & 53,000 & 114 & \nodata & 975 & Akerman  &
073.D-0024(A) & 2004-09-01\\
HI.20050617.48772.fits & HIRES & 3930$-$6895 & 103,000 & 106 & 197 & 1200 &
M\'{e}lendez & C01H & 2005-06-17\\
HI.20050617.50045.fits & HIRES & 3930$-$6895 & 103,000 & 103 & 179 & 1200 &
M\'{e}lendez & C01H & 2005-06-17\\
HI.20050617.51319.fits & HIRES & 3930$-$6895 & 103,000 & 108 & 164 & 1200 &
M\'{e}lendez & C01H & 2005-06-17\\
\hline
\hline
Combined Spectrum & \nodata & 3750$-$6895 & 51,700 & 356 & 523 &
\nodata & \nodata & This Study & \nodata\\
\hline
\hline
\nodata & AAT-echelle & 3700$-$4700 & 42,000 & 102\tablenotemark{c} &
\nodata & \nodata & \nodata & NRB01 & \nodata\\
\nodata & Subaru-HRS & 4030$-$6780 & 55,000 & \nodata &
249\tablenotemark{d} & 3600 & \nodata & Ishigaki et al.\ (2010) &
\nodata\\
\nodata & Magellan-MIKE & 3350$-$9100 & 30,000 & 51 & 90 &
\nodata & \nodata & Frebel et al.\ (2013) & \nodata\\
\enddata
\tablenotetext{a}{S/N per pixel.}
\tablenotetext{b}{This column Proposal ID of archival spectra used in this
  work or else reference to literature data compared to in the text.}
\tablenotetext{c}{Square root of number of photons per pixel at
  4300\AA\ as described in NRB01.}
\tablenotetext{d}{S/N per resolution element measured at 5800\AA.}
\end{deluxetable*}
\clearpage
\end{landscape}
\clearpage

\begin{deluxetable*}{lccrccrc}
\tablewidth{0pt}
\tabletypesize{\scriptsize}
\tablecaption{\label{tab_ews} Equivalent Widths of CD~$-$24$^{o}$17504}
\tablehead{
\colhead{Species} & \colhead{$\lambda$ (\AA)} & \colhead{E.P.} &
\colhead{log{\it gf}} & \colhead{EW (m\AA)} & \colhead{$\Delta$EW (m\AA)} &
\colhead{log$\epsilon$(X)\tablenotemark{a}} &
\colhead{ULflag\tablenotemark{b}}
}
\startdata
%El.      Wavelength  EP        loggf     EW      EWerr    abund  ulflag 
Li~I  &  6707.800  & 0.00  &    0.170  &  18.5  &  2.0 &  1.99 & 0 \\   
CH    &  4313      & \nodata & \nodata &   syn  &  \nodata &  6.15 & 0 \\
CH    &  4323      & \nodata & \nodata &   syn  &  \nodata &  6.08 & 0\\
Na~I  &  5889.950  & 0.00  &    0.108  &  26.6  &  1.7 &  2.61 & 0 \\   
Na~I  &  5895.924  & 0.00  & $-$0.194  &  14.6  &  1.5 &  2.59 & 0 \\   
Mg~I  &  4057.505  & 4.35  & $-$0.890  &   2.1  &  0.5 &  4.43 & 0 \\
Mg~I  &  4167.271  & 4.35  & $-$0.710  &   4.2  &  0.8 &  4.55 & 0 \\
Mg~I  &  4702.990  & 4.33  & $-$0.380  &   8.8  &  1.0 &  4.53 & 0 \\
Mg~I  &  5172.684  & 2.71  & $-$0.450  &  75.2  &  1.2 &  4.55 & 0 \\
Mg~I  &  5183.604  & 2.72  & $-$0.239  &  89.4  &  1.6 &  4.60 & 0 \\
Mg~I  &  5528.405  & 4.34  & $-$0.498  &   7.4  &  0.4 &  4.56 & 0 \\
Al~I  &  3944.010  & 0.00  & $-$0.620  &   syn  &  \nodata &  2.47 & 0 \\
Al~I  &  3961.520  & 0.01  & $-$0.340  &  24.6  &  1.1 &  2.35 & 0 \\
Al~I  &  3961.520  & 0.01  & $-$0.340  &   syn  &  \nodata &  2.40 & 0 \\
Si~I  &  3905.523  & 1.91  & $-$1.092  &  52.6  &  1.3 &  4.19 & 0 \\
Si~I  &  3905.523  & 1.91  & $-$1.092  &   syn  &  \nodata &  4.20 & 0 \\
Ca~I  &  4226.730  & 0.00  &    0.244  &  74.9  &  1.5 &  3.14 & 0 \\
Ca~I  &  4283.010  & 1.89  & $-$0.224  &   3.7  &  0.8 &  3.24 & 0 \\
Ca~I  &  4318.650  & 1.89  & $-$0.210  &   3.2  &  0.4 &  3.17 & 0 \\
Ca~I  &  4425.440  & 1.88  & $-$0.358  &   3.0  &  0.6 &  3.27 & 0 \\
Ca~I  &  4434.960  & 1.89  & $-$0.010  &   4.9  &  0.5 &  3.15 & 0 \\
Ca~I  &  4435.690  & 1.89  & $-$0.519  &   2.2  &  0.7 &  3.30 & 0 \\
Ca~I  &  4454.780  & 1.90  &    0.260  &   8.7  &  0.6 &  3.16 & 0 \\
Ca~I  &  5588.760  & 2.52  &    0.210  &   3.3  &  1.5 &  3.29 & 0 \\
Ca~I  &  6122.220  & 1.89  & $-$0.315  &   2.6  &  0.6 &  3.12 & 0 \\
Ca~I  &  6162.170  & 1.90  & $-$0.089  &   4.0  &  0.5 &  3.09 & 0 \\
Ca~I  &  6439.070  & 2.52  &    0.470  &   3.3  &  0.6 &  3.00 & 0 \\
Sc~II &  4246.820  & 0.32  &    0.240  &   syn  &  \nodata & $-$0.01 & 0 \\
Sc~II &  4314.083  & 0.62  & $-$0.100  &   syn  &  \nodata & $-$0.06 & 0 \\
Sc~II &  4324.998  & 0.59  & $-$0.440  &   syn  &  \nodata & $-$0.22 & 0 \\
Sc~II &  4400.389  & 0.61  & $-$0.540  &   syn  &  \nodata & $-$0.01 & 0 \\
Sc~II &  4415.540  & 0.59  & $-$0.670  &   syn  &  \nodata &  0.07 & 0 \\
Ti~I  &  3989.760  & 0.02  & $-$0.062  &   3.8  &  0.8 &  2.13 & 0 \\   
Ti~I  &  3998.640  & 0.05  &    0.010  &   5.2  &  0.7 &  2.23 & 0 \\   
Ti~I  &  4533.249  & 0.85  &    0.532  &   3.6  &  0.3 &  2.25 & 0 \\   
Ti~I  &  4534.780  & 0.84  &    0.336  &   2.1  &  0.2 &  2.19 & 0 \\   
Ti~II &  3913.461  & 1.12  & $-$0.420  &  22.7  &  1.0 &  1.94 & 0 \\   
Ti~II &  4012.396  & 0.57  & $-$1.750  &   4.2  &  0.6 &  1.89 & 0 \\
Ti~II &  4163.634  & 2.59  & $-$0.400  &   1.8  &  0.3 &  2.03 & 0 \\
Ti~II &  4290.219  & 1.16  & $-$0.930  &   7.6  &  0.4 &  1.88 & 0 \\
Ti~II &  4300.049  & 1.18  & $-$0.490  &  16.4  &  1.0 &  1.85 & 0 \\
Ti~II &  4395.031  & 1.08  & $-$0.540  &  19.4  &  1.1 &  1.89 & 0 \\
Ti~II &  4399.765  & 1.24  & $-$1.190  &   3.5  &  0.3 &  1.85 & 0 \\
Ti~II &  4417.714  & 1.17  & $-$1.190  &   3.6  &  0.4 &  1.79 & 0 \\
Ti~II &  4443.801  & 1.08  & $-$0.720  &  14.9  &  0.6 &  1.92 & 0 \\
Ti~II &  4450.482  & 1.08  & $-$1.520  &   2.9  &  0.6 &  1.95 & 0 \\
Ti~II &  4468.517  & 1.13  & $-$0.600  &  15.0  &  0.8 &  1.85 & 0 \\
Ti~II &  4501.270  & 1.12  & $-$0.770  &  12.7  &  0.7 &  1.92 & 0 \\
Ti~II &  4533.960  & 1.24  & $-$0.530  &  13.4  &  0.5 &  1.82 & 0 \\
Ti~II &  4563.770  & 1.22  & $-$0.960  &   9.1  &  0.5 &  2.03 & 0 \\
Ti~II &  4571.971  & 1.57  & $-$0.320  &  12.7  &  0.5 &  1.88 & 0 \\
Ti~II &  5188.687  & 1.58  & $-$1.050  &   2.2  &  0.4 &  1.79 & 0 \\
Cr~I  &  4254.332  & 0.00  & $-$0.114  &  19.3  &  0.5 &  2.13 & 0 \\
Cr~I  &  4274.800  & 0.00  & $-$0.220  &  16.6  &  0.7 &  2.16 & 0 \\
Cr~I  &  4289.720  & 0.00  & $-$0.370  &  13.8  &  0.6 &  2.20 & 0 \\
Cr~I  &  5206.040  & 0.94  &    0.020  &   7.0  &  0.7 &  2.30 & 0 \\
Cr~I  &  5208.419  & 0.94  &    0.160  &  10.2  &  0.6 &  2.35 & 0 \\
Mn~I  &  4030.753  & 0.00  & $-$0.480  &   syn  &  \nodata &  2.00 & 0 \\
Mn~I  &  4033.062  & 0.00  & $-$0.618  &   syn  &  \nodata &  2.04 & 0 \\
Mn~I  &  4034.483  & 0.00  & $-$0.811  &   syn  &  \nodata &  2.05 & 0 \\
Fe~I  &  3786.677  & 1.01  & $-$2.185  &   3.1  &  0.6 &  4.03 & 0 \\   
Fe~I  &  3787.880  & 1.01  & $-$0.838  &  38.1  &  0.6 &  4.05 & 0 \\
Fe~I  &  3805.343  & 3.30  &    0.313  &   7.0  &  0.5 &  4.00 & 0 \\
Fe~I  &  3815.840  & 1.48  &    0.237  &  64.5  &  1.1 &  4.04 & 0 \\   
Fe~I  &  3820.425  & 0.86  &    0.157  &  82.0  &  0.8 &  4.12 & 0 \\   
Fe~I  &  3824.444  & 0.00  & $-$1.360  &  61.3  &  0.6 &  4.23 & 0 \\   
Fe~I  &  3825.881  & 0.91  & $-$0.024  &  70.1  &  0.7 &  4.00 & 0 \\   
Fe~I  &  3827.823  & 1.56  &    0.094  &  50.1  &  0.8 &  3.91 & 0 \\   
Fe~I  &  3839.256  & 3.05  & $-$0.330  &   2.6  &  0.4 &  3.96 & 0 \\   
Fe~I  &  3840.438  & 0.99  & $-$0.497  &  44.6  &  0.6 &  3.84 & 0 \\   
Fe~I  &  3841.048  & 1.61  & $-$0.044  &  41.1  &  0.7 &  3.87 & 0 \\   
Fe~I  &  3846.800  & 3.25  & $-$0.020  &   3.8  &  0.8 &  4.00 & 0 \\
Fe~I  &  3849.967  & 1.01  & $-$0.863  &  38.1  &  0.7 &  4.06 & 0 \\
Fe~I  &  3850.818  & 0.99  & $-$1.745  &  10.2  &  0.8 &  4.12 & 0 \\
Fe~I  &  3852.573  & 2.18  & $-$1.180  &   2.7  &  0.5 &  4.01 & 0 \\
Fe~I  &  3856.372  & 0.05  & $-$1.280  &  64.3  &  0.8 &  4.28 & 0 \\   
Fe~I  &  3859.911  & 0.00  & $-$0.710  &  86.3  &  0.6 &  4.32 & 0 \\   
Fe~I  &  3865.523  & 1.01  & $-$0.950  &  33.9  &  0.7 &  4.06 & 0 \\   
Fe~I  &  3867.216  & 3.02  & $-$0.450  &   3.1  &  0.5 &  4.13 & 0 \\
Fe~I  &  3878.018  & 0.96  & $-$0.896  &  37.6  &  0.5 &  4.04 & 0 \\
Fe~I  &  3878.573  & 0.09  & $-$1.380  &  57.1  &  0.7 &  4.20 & 0 \\
Fe~I  &  3895.656  & 0.11  & $-$1.668  &  38.2  &  0.7 &  4.03 & 0 \\
Fe~I  &  3899.707  & 0.09  & $-$1.515  &  48.9  &  1.0 &  4.11 & 0 \\
Fe~I  &  3902.946  & 1.56  & $-$0.442  &  33.1  &  0.7 &  4.03 & 0 \\
Fe~I  &  3917.181  & 0.99  & $-$2.155  &   4.3  &  0.5 &  4.11 & 0 \\
Fe~I  &  3920.258  & 0.12  & $-$1.734  &  41.0  &  0.8 &  4.16 & 0 \\
Fe~I  &  3922.912  & 0.05  & $-$1.626  &  48.9  &  0.9 &  4.18 & 0 \\   
Fe~I  &  3977.741  & 2.20  & $-$1.120  &   3.5  &  0.6 &  4.10 & 0 \\   
Fe~I  &  4005.242  & 1.56  & $-$0.583  &  29.8  &  0.8 &  4.08 & 0 \\
Fe~I  &  4021.866  & 2.76  & $-$0.730  &   2.7  &  0.6 &  4.10 & 0 \\
Fe~I  &  4045.812  & 1.49  &    0.284  &  69.4  &  0.8 &  4.15 & 0 \\   
Fe~I  &  4062.441  & 2.85  & $-$0.860  &   2.0  &  0.5 &  4.17 & 0 \\   
Fe~I  &  4063.594  & 1.56  &    0.062  &  57.2  &  0.7 &  4.10 & 0 \\   
Fe~I  &  4067.978  & 3.21  & $-$0.470  &   1.6  &  0.4 &  4.01 & 0 \\   
Fe~I  &  4071.738  & 1.61  & $-$0.008  &  52.3  &  0.7 &  4.08 & 0 \\   
Fe~I  &  4076.629  & 3.21  & $-$0.370  &   1.6  &  0.3 &  3.91 & 0 \\   
Fe~I  &  4132.058  & 1.61  & $-$0.675  &  25.3  &  0.9 &  4.11 & 0 \\
Fe~I  &  4134.678  & 2.83  & $-$0.649  &   3.6  &  0.5 &  4.20 & 0 \\
Fe~I  &  4136.998  & 3.42  & $-$0.450  &   1.4  &  0.5 &  4.12 & 0 \\
Fe~I  &  4143.414  & 3.05  & $-$0.200  &   5.3  &  0.5 &  4.13 & 0 \\
Fe~I  &  4143.868  & 1.56  & $-$0.511  &  33.6  &  0.7 &  4.09 & 0 \\
Fe~I  &  4147.669  & 1.48  & $-$2.071  &   1.8  &  0.4 &  4.09 & 0 \\
Fe~I  &  4153.899  & 3.40  & $-$0.320  &   1.8  &  0.5 &  4.08 & 0 \\
Fe~I  &  4154.498  & 2.83  & $-$0.688  &   2.0  &  0.5 &  3.98 & 0 \\
Fe~I  &  4154.805  & 3.37  & $-$0.400  &   1.5  &  0.6 &  4.05 & 0 \\
Fe~I  &  4156.799  & 2.83  & $-$0.808  &   2.3  &  0.7 &  4.16 & 0 \\
Fe~I  &  4157.780  & 3.42  & $-$0.403  &   2.6  &  0.5 &  4.34 & 0 \\
Fe~I  &  4174.913  & 0.91  & $-$2.938  &   1.3  &  0.5 &  4.26 & 0 \\
Fe~I  &  4181.755  & 2.83  & $-$0.371  &   6.4  &  0.7 &  4.19 & 0 \\
Fe~I  &  4184.892  & 2.83  & $-$0.869  &   2.0  &  0.4 &  4.16 & 0 \\
Fe~I  &  4187.039  & 2.45  & $-$0.514  &   7.9  &  0.5 &  4.08 & 0 \\
Fe~I  &  4187.795  & 2.42  & $-$0.510  &   7.8  &  0.6 &  4.04 & 0 \\
Fe~I  &  4191.430  & 2.47  & $-$0.666  &   6.1  &  0.3 &  4.13 & 0 \\
Fe~I  &  4195.329  & 3.33  & $-$0.492  &   1.8  &  0.5 &  4.18 & 0 \\
Fe~I  &  4199.095  & 3.05  &    0.156  &   9.5  &  0.8 &  4.05 & 0 \\
Fe~I  &  4202.029  & 1.49  & $-$0.689  &  29.6  &  0.9 &  4.10 & 0 \\
Fe~I  &  4216.184  & 0.00  & $-$3.357  &   2.8  &  0.4 &  4.17 & 0 \\
Fe~I  &  4222.213  & 2.45  & $-$0.914  &   3.2  &  0.6 &  4.06 & 0 \\
Fe~I  &  4227.427  & 3.33  &    0.266  &   8.2  &  0.7 &  4.12 & 0 \\
Fe~I  &  4233.603  & 2.48  & $-$0.579  &   6.5  &  0.7 &  4.08 & 0 \\
Fe~I  &  4238.810  & 3.40  & $-$0.233  &   2.1  &  0.7 &  4.05 & 0 \\
Fe~I  &  4247.426  & 3.37  & $-$0.240  &   2.7  &  0.9 &  4.15 & 0 \\
Fe~I  &  4250.119  & 2.47  & $-$0.380  &   9.3  &  0.7 &  4.04 & 0 \\   
Fe~I  &  4250.787  & 1.56  & $-$0.713  &  24.9  &  0.5 &  4.08 & 0 \\   
Fe~I  &  4260.474  & 2.40  &    0.077  &  24.2  &  0.7 &  4.03 & 0 \\
Fe~I  &  4271.154  & 2.45  & $-$0.337  &  11.6  &  0.8 &  4.09 & 0 \\
Fe~I  &  4271.760  & 1.49  & $-$0.173  &  53.8  &  0.6 &  4.15 & 0 \\
Fe~I  &  4282.403  & 2.18  & $-$0.779  &   7.3  &  0.5 &  4.06 & 0 \\
Fe~I  &  4325.762  & 1.61  &    0.006  &  53.7  &  0.3 &  4.07 & 0 \\
Fe~I  &  4352.735  & 2.22  & $-$1.290  &   2.2  &  0.9 &  4.05 & 0 \\
Fe~I  &  4375.930  & 0.00  & $-$3.005  &   5.9  &  0.5 &  4.13 & 0 \\
Fe~I  &  4383.545  & 1.48  &    0.200  &  69.4  &  0.7 &  4.17 & 0 \\
Fe~I  &  4404.750  & 1.56  & $-$0.147  &  52.4  &  0.7 &  4.14 & 0 \\
Fe~I  &  4415.122  & 1.61  & $-$0.621  &  29.9  &  0.5 &  4.13 & 0 \\   
Fe~I  &  4427.310  & 0.05  & $-$2.924  &   6.7  &  0.6 &  4.16 & 0 \\
Fe~I  &  4442.339  & 2.20  & $-$1.228  &   3.1  &  0.6 &  4.16 & 0 \\
Fe~I  &  4447.717  & 2.22  & $-$1.339  &   2.5  &  0.4 &  4.15 & 0 \\
Fe~I  &  4459.118  & 2.18  & $-$1.279  &   3.3  &  0.6 &  4.18 & 0 \\
Fe~I  &  4461.653  & 0.09  & $-$3.194  &   4.0  &  0.7 &  4.22 & 0 \\
Fe~I  &  4466.552  & 2.83  & $-$0.600  &   2.5  &  1.0 &  3.95 & 0 \\
Fe~I  &  4476.019  & 2.85  & $-$0.820  &   2.9  &  0.6 &  4.27 & 0 \\
Fe~I  &  4528.614  & 2.18  & $-$0.822  &   7.6  &  0.6 &  4.10 & 0 \\
Fe~I  &  4531.148  & 1.48  & $-$2.101  &   1.7  &  0.5 &  4.05 & 0 \\
Fe~I  &  4602.941  & 1.49  & $-$2.208  &   1.2  &  0.6 &  4.00 & 0 \\
Fe~I  &  4871.318  & 2.87  & $-$0.362  &   4.3  &  0.6 &  3.99 & 0 \\
Fe~I  &  4890.755  & 2.88  & $-$0.394  &   4.3  &  0.5 &  4.03 & 0 \\
Fe~I  &  4891.492  & 2.85  & $-$0.111  &   9.1  &  0.4 &  4.07 & 0 \\
Fe~I  &  4903.310  & 2.88  & $-$0.926  &   1.3  &  0.5 &  4.03 & 0 \\
Fe~I  &  4918.994  & 2.85  & $-$0.342  &   5.1  &  0.6 &  4.03 & 0 \\
Fe~I  &  4920.503  & 2.83  &    0.068  &  13.1  &  0.7 &  4.06 & 0 \\   
Fe~I  &  5012.068  & 0.86  & $-$2.642  &   1.9  &  0.9 &  4.03 & 0 \\
Fe~I  &  5083.339  & 0.96  & $-$2.842  &   1.7  &  0.7 &  4.27 & 0 \\
Fe~I  &  5171.596  & 1.49  & $-$1.721  &   4.3  &  0.7 &  4.05 & 0 \\
Fe~I  &  5192.344  & 3.00  & $-$0.421  &   3.2  &  0.5 &  4.02 & 0 \\
Fe~I  &  5194.942  & 1.56  & $-$2.021  &   1.9  &  1.2 &  4.05 & 0 \\
Fe~I  &  5232.940  & 2.94  & $-$0.057  &   8.8  &  0.6 &  4.07 & 0 \\   
Fe~I  &  5266.555  & 3.00  & $-$0.385  &   3.5  &  0.7 &  4.03 & 0 \\
Fe~I  &  5269.537  & 0.86  & $-$1.333  &  34.4  &  1.0 &  4.20 & 0 \\   
Fe~I  &  5328.039  & 0.92  & $-$1.466  &  13.5  &  0.6 &  3.81 & 0 \\
Fe~I  &  5371.489  & 0.96  & $-$1.644  &   8.9  &  0.7 &  3.81 & 0 \\
Fe~I  &  5383.369  & 4.31  &    0.645  &   2.7  &  0.7 &  4.07 & 0 \\   
Fe~I  &  5397.128  & 0.92  & $-$1.982  &   9.8  &  1.0 &  4.16 & 0 \\
Fe~I  &  5405.775  & 0.99  & $-$1.852  &  11.0  &  0.6 &  4.15 & 0 \\
Fe~I  &  5415.199  & 4.39  &    0.643  &   2.5  &  0.5 &  4.11 & 0 \\
Fe~I  &  5424.068  & 4.32  &    0.520  &   3.5  &  1.1 &  4.32 & 0 \\
Fe~I  &  5429.696  & 0.96  & $-$1.881  &  11.1  &  1.0 &  4.16 & 0 \\
Fe~I  &  5434.524  & 1.01  & $-$2.126  &   5.6  &  1.1 &  4.12 & 0 \\   
Fe~I  &  5446.917  & 0.99  & $-$1.910  &  10.7  &  1.3 &  4.20 & 0 \\
Fe~I  &  5455.609  & 1.01  & $-$2.090  &   7.0  &  0.7 &  4.19 & 0 \\
Fe~I  &  5497.516  & 1.01  & $-$2.825  &   1.0  &  0.5 &  4.05 & 0 \\
Fe~I  &  5506.779  & 0.99  & $-$2.789  &   1.6  &  0.3 &  4.20 & 0 \\
Fe~I  &  5586.756  & 3.37  & $-$0.144  &   3.2  &  0.8 &  4.07 & 0 \\   
Fe~I  &  5615.644  & 3.33  &    0.050  &   4.7  &  0.8 &  4.02 & 0 \\
Fe~II &  4233.170  & 2.58  & $-$1.970  &   6.4  &  1.3 &  4.08 & 0 \\
Fe~II &  4522.630  & 2.84  & $-$2.250  &   2.9  &  0.7 &  4.21 & 0 \\
Fe~II &  4583.840  & 2.81  & $-$1.930  &   5.0  &  0.7 &  4.11 & 0 \\
Fe~II &  4923.930  & 2.89  & $-$1.320  &  12.4  &  0.8 &  4.00 & 0 \\
Fe~II &  5018.450  & 2.89  & $-$1.220  &  16.4  &  1.1 &  4.04 & 0 \\
Co~I  &  3845.468  & 0.92  &    0.010  &   7.6  &  1.0 &  2.01 & 0 \\
Co~I  &  3873.120  & 0.43  & $-$0.660  &   9.0  &  0.9 &  2.31 & 0 \\
Co~I  &  3995.306  & 0.92  & $-$0.220  &   7.2  &  0.8 &  2.20 & 0 \\
Co~I  &  4121.318  & 0.92  & $-$0.320  &   4.8  &  0.6 &  2.10 & 0 \\
Ni~I  &  3783.520  & 0.42  & $-$1.420  &  16.2  &  1.1 &  3.20 & 0 \\
Ni~I  &  3807.140  & 0.42  & $-$1.220  &  19.6  &  0.7 &  3.09 & 0 \\
Ni~I  &  3858.301  & 0.42  & $-$0.951  &  32.6  &  0.8 &  3.15 & 0 \\
Ni~I  &  5476.900  & 1.83  & $-$0.890  &   4.3  &  0.4 &  3.19 & 0 \\
Zn~I  &  4810.528  & 4.08  & $-$0.137  &   3.1  &  1.0 &  2.15 & 0 \\
Sr~II &  4077.714  & 0.00  &    0.150  &   5.5  & 0.6 & $-$1.81 & 0 \\   
Sr~II &  4077.714  & 0.00  &    0.150  &   syn  & \nodata & $-$1.82 & 0 \\   
Sr~II &  4215.524  & 0.00  & $-$0.180  &   3.1  & 0.7 & $-$1.75 & 0 \\   
Sr~II &  4215.524  & 0.00  & $-$0.180  &   syn  & \nodata & $-$1.80 & 0 \\
Y~II  &  3788.694  & 0.10  & $-$0.140  &$<$1.0  & \nodata & $< -$1.36 & 1 \\
Zr~II &  4208.977  & 0.71  & $-$0.460  &$<$1.0  & \nodata & $<  $0.08 & 1 \\
Ba~II &  4554.033  & 0.00  &    0.163  &$<$1.0  & \nodata & $< -$2.28 & 1 \\
La~II &  4123.220  & 0.32  &    0.130  &$<$1.0  & \nodata & $< -$1.08 & 1 \\
Eu~II &  4129.700  & 0.00  &    0.220  &$<$1.0  & \nodata & $< -$1.73 & 1 \\
\enddata
\tablenotetext{a}{\,LTE abundance}
\tablenotetext{b}{Upper limit flag: 1 = yes, 0 = no}
\end{deluxetable*}

\clearpage

\begin{deluxetable}{lcclcl}
\tablecaption{Atmospheric Parameters for \cd24\ in this study and in
  the literature\label{tab_params}}
\tablewidth{0pt}
\setlength{\tabcolsep}{0.01in}
\tabletypesize{\scriptsize}
\tablehead{
\colhead{\teff} &
\colhead{\logg} &
\colhead{\vt} & 
\colhead{\feh} &
\colhead{} &
\colhead{} \\
\colhead{(K)} &
\colhead{(dex)} &
\colhead{(km s$^{-1}$)} &
\colhead{(dex)} &
\colhead{Method} &
\colhead{Ref.} } 
\startdata
6228 & 3.90 & 1.25 & $-$3.41 & spec\tablenotemark{a}, LTE & This Study \\
6228 & 4.23 & 1.00 & $-$3.29 & spec\tablenotemark{a}, NLTE & This Study \\
6259 & 3.65 & 1.40 & $-$3.23 & spec\tablenotemark{a} & \citet{teff_calib} \\
6236 & 3.70 & 1.60 & $-$3.23 & phot & \citet{yong13_II} \\
5821 & 3.50 & 1.22 & $-$3.66 & spec & \citet{ishigaki2010} \\
6456 & 4.74 & 1.50 & $-$3.20 & comb & \citet{ishigaki2012} \\
6451 & 4.13 & \nodata & $-$3.34 & phot & \citet{melendez2010} \\
6180 & 4.40 & 1.50 & $-$3.40 & balm & \citet{aoki2009li} \\
6070 & 3.57 & 1.30 & $-$3.35 & comb & \citet{hosford09} \\
5942 & 4.05 & 1.50 & $-$3.42 & spec & \citet{rich09} \\
6338 & 4.32 & 1.50 & $-$3.21 & balm & \citet{nissen2007} \\
6070 & 4.20 & 1.80 & $-$3.45 & phot & \citet{arnone} \\
6212 & 4.13 & 1.00 & $-$2.99 & phot & \citet{bihain2004} \\
6212 & 4.13 & \nodata & $-$3.32 & phot & \citet{israelian2001} \\
6070 & 3.60 & 1.40 & $-$3.37 & comb & \citet{Norrisetal:2001} \\
6300 & 4.50 & 1.00 & $-$3.30 & phot & \citet{primas} \\
6100 & 4.00 & 1.50 & $-$3.70 & balm & \citet{spite1996li} \\
\enddata
\tablecomments{Methods for determining stellar parameters range from
  classical spectroscopic methods (``spec''), use of color-temperature
  relations (``phot''), or fitting of Balmer line absorption wings
  (``balm'').  In the latter two methods, \logg\ is often determined by
  comparison to theoretical isochrones, but in some cases, is
  determined by ionization balance.  These cases are noted as
  ``comb'', for combination of methods.}
\tablenotetext{a}{Parameters determined spectroscopically, but with
correction applied to \teff.  See text for more information.}
\end{deluxetable}

\begin{deluxetable}{lcccccc}
\tabletypesize{\scriptsize}
\tablewidth{0pt}
\tablecaption{Uncertainties in Stellar Parameters Due to Errors in
  Fe\,I, Fe\,II EWs\label{par_unc}}
\tablehead{
\colhead{} & \colhead{$\Delta$\teff} & \colhead{$\Delta$\logg} &
\colhead{$\Delta$\vt} & \colhead{$\Delta$\feh} & 
\colhead{} & \colhead{} \\
\colhead{Run} & \colhead{(K)} & \colhead{(dex)} & \colhead{(\kms)} &
\colhead{(dex)} & \colhead{\#\ Fe\,I} & \colhead{\#\ Fe\,II} 
}
\startdata
01 & $-$68 & $-$0.18 & $+$0.00 & $-$0.04 & 112 & 5 \\
02 & $+$76 & $-$0.02 & $+$0.06 & $+$0.07 & 112 & 5 \\
03 & $-$67 & $-$0.24 & $+$0.06 & $-$0.06 & 109 & 5 \\
04 & $+$13 & $+$0.02 & $+$0.05 & $+$0.01 & 110 & 5 \\
05 & $-$51 & $-$0.25 & $+$0.00 & $-$0.03 & 112 & 5 \\
06 & $+$123 & $+$0.48 & $+$0.04 & $+$0.10 & 110 & 5 \\
07 & $+$37 & $+$0.09 & $+$0.08 & $+$0.02 & 112 & 5 \\
08 & $-$71 & $-$0.05 & $-$0.09 & $-$0.04 & 110 & 5 \\
09 & $-$6 & $-$0.03 & $+$0.08 & $-$0.01 & 110 & 5 \\
10 & $-$74 & $-$0.13 & $-$0.02 & $-$0.05 & 113 & 5 \\
\hline
ave. & 59 & 0.15 & 0.05 & 0.04 & \nodata & \nodata \\
$\sigma$ & 34 & 0.15 & 0.03 & 0.03 & \nodata & \nodata\\
\enddata
\end{deluxetable}

\clearpage

\begin{deluxetable}{lcccccc}
\tabletypesize{\scriptsize}
\tablewidth{0pt}
\tablecaption{Log($\epsilon$) Abundance Uncertainties due to Atmospheric Parameters\label{tab_unc}}
\tablehead{
\colhead{} &
\colhead{} &
\colhead{$\Delta$\teff(K)} &
\colhead{$\Delta$\logg} &
\colhead{$\Delta$\vt} & 
\colhead{$\Delta$[M/H]} & \colhead{}\\
\colhead{Species} & \colhead{$\sigma$(E.W.)\tablenotemark{a}} &
\colhead{$+$84 K} &
\colhead{$+$0.34 dex} &
\colhead{$+$0.11 km s$^{-1}$} & \colhead{$+$0.11 dex} & \colhead{Total}
}
\startdata
 Li\,I & 0.03\tablenotemark{b}  & $+$0.06 & $+$0.00 & $+$0.00 & $+$0.00 & 0.07\\
 CH    & 0.10\tablenotemark{b} & $+$0.15 & $-$0.15 & $+$0.00 & $+$0.00 & 0.23 \\
 Na\,I & 0.02 & $+$0.06 & $-$0.01 & $+$0.00 & $+$0.00 & 0.06 \\
 Mg\,I & 0.07 & $+$0.04 & $-$0.03 & $-$0.01 & $+$0.00 & 0.09 \\
 Al\,I & 0.09 & $+$0.05 & $-$0.01 & $-$0.01 & $+$0.00 & 0.10\\
 Si\,I & 0.15\tablenotemark{c} & $+$0.05 & $-$0.01 & $-$0.02 & $-$0.01 & 0.16\\
 Ca\,I & 0.08 & $+$0.05 & $-$0.01 & $+$0.00 & $+$0.01 & 0.10\\
 Sc\,II & 0.10 & $+$0.04 & $+$0.11 & $+$0.00 & $+$0.00 & 0.15 \\
 Ti\,I & 0.05 & $+$0.07 & $+$0.00 & $+$0.00 & $+$0.01 & 0.09 \\
 Ti\,II & 0.11 & $+$0.04 & $+$0.12 & $+$0.00 & $+$0.00 & 0.18\\
 Cr\,I & 0.09 & $+$0.07 & $-$0.01 & $+$0.00 & $+$0.00 & 0.11 \\
 Mn\,I & 0.20 & $+$0.08 & $+$0.00 & $+$0.00 & $+$0.01 & 0.22 \\
 Fe\,I & 0.10 & $+$0.06 & $-$0.01 & $-$0.01 & $+$0.00
 & 0.12 \\
 Fe\,II & 0.07 & $+$0.01 & $+$0.12 & $+$0.00 &
 $+$0.00 & 0.14 \\
 Co\,I & 0.11 & $+$0.07 & $+$0.00 & $+$0.00 & $+$0.00 & 0.13\\
 Ni\,I & 0.06 & $+$0.08 & $+$0.01 & $+$0.00 & $+$0.01 & 0.10 \\
 Zn\,I & 0.15\tablenotemark{c} & $+$0.04 & $+$0.03 & $+$0.00 & $+$0.00 & 0.16\\
 Sr\,II & 0.10 & $+$0.05 & $+$0.11 & $+$0.00 & $+$0.00 & 0.16 \\
 Y\,II & 0.18\tablenotemark{d} & $+$0.05 & $+$0.11 & $+$0.00 & $+$0.00 & 0.22 \\
 Zr\,II & 0.18\tablenotemark{d} & $+$0.04 & $+$0.11 & $+$0.00 & $+$0.00 & 0.21\\
 Ba\,II & 0.16\tablenotemark{d} & $+$0.05 & $+$0.10 &$+$0.10 & $+$0.00 & 0.20 \\
 La\,II & 0.17\tablenotemark{d} & $+$0.03 & $+$0.11 & $-$0.01 & $-$0.01 & 0.21\\
 Eu\,II & 0.16\tablenotemark{d} & $+$0.05 & $+$0.11 & $+$0.00 & $+$0.00 & 0.20\\
\enddata
\tablenotetext{a}{The maximum of the standard deviation of individual
  line element abundances or the abundance sensitivity to EW uncertainties.}
\tablenotetext{b}{Sensitivity of abundance to continuum placement.}
\tablenotetext{c}{Value given to measures based on one line.}
\tablenotetext{d}{EW uncertainty set to 0.5 m\AA.  See text for more information.}
\end{deluxetable}

\clearpage

\begin{deluxetable}{lccrccc}
\tablewidth{0pt}
\tabletypesize{\scriptsize}
\tablecaption{\label{tab_lte_abund} Element Abundances for
  CD~$-$24$^{o}$17504 Based on LTE Stellar Parameters}
\tablehead{
\colhead{Species} & \colhead{\#\ lines} & \colhead{log$\epsilon$(X)} &
\colhead{$\sigma$} & \colhead{$\rm[X/H]$} & \colhead{$\rm[X/Fe]$} &
\colhead{$\sigma$/$\sqrt(N)$}
}
\startdata
Li\,I  &   1 &      1.99 & 0.10 &   \nodata & \nodata & \nodata \\
1D CH     &   2 &      6.12 & 0.05 &   $-$2.31 &   $+$1.10 & 0.04 \\
3D CH\tablenotemark{a}     &   2 & 5.52 & 0.05 & $-$2.91 & $+$0.50 & 0.01\\
LTE C\,I\tablenotemark{b} & 2 & 5.71 & 0.05 & $-$2.72 & $+$0.69 & 0.03\\
NLTE C\,I\tablenotemark{c} & 2 & 5.45 & 0.05 & $-$2.98 & $+$0.43 & 0.03  \\
LTE O\,I\tablenotemark{b} & 1 & 6.12 & \nodata & $-$2.57 & $+$0.84 & \nodata \\
NLTE O\,I\tablenotemark{d} & 1 & 5.70 & \nodata & $-$2.99 & $+$0.42 &
\nodata \\
Na\,I  &   2 &      2.60 & 0.01 &   $-$3.64 &   $-$0.23 & 0.01 \\
Mg\,I  &   6 &      4.54 & 0.05 &   $-$3.06 &   $+$0.34 & 0.02 \\
Al\,I  &   2 &      2.44 & 0.05 &   $-$4.02 &   $-$0.61 & 0.04 \\
Si\,I  &   1 &      4.20 & 0.15 &   $-$3.31 &   $+$0.10 & 0.15 \\
Ca\,I  &  11 &      3.18 & 0.09 &   $-$3.16 &   $+$0.24 & 0.03 \\
Sc\,II &   5 &   $-$0.05 & 0.11 &   $-$3.20 &   $+$0.21 & 0.05 \\
Ti\,I  &   4 &      2.20 & 0.05 &   $-$2.75 &   $+$0.66 & 0.03 \\
Ti\,II &  16 &      1.89 & 0.07 &   $-$3.06 &   $+$0.35 & 0.02 \\
Cr\,I  &   5 &      2.23 & 0.08 &   $-$3.41 &   $-$0.01 & 0.04 \\
Mn\,I  &   3 &      2.03 & 0.03 &   $-$3.40 &   $+$0.01 & 0.02 \\
Fe\,I  & 113 &      4.09 & 0.10 &   $-$3.41 &   \nodata & 0.01 \\
Fe\,II &   5 &      4.09 & 0.07 &   $-$3.41 &   \nodata & 0.03 \\
Co\,I  &   4 &      2.16 & 0.11 &   $-$2.83 &   $+$0.57 & 0.06 \\
Ni\,I  &   4 &      3.16 & 0.04 &   $-$3.06 &   $+$0.34 & 0.03 \\
Zn\,I  &   1 &      2.15 & 0.15 &   $-$2.41 &   $+$1.00 & 0.15 \\
Sr\,II &   2 &   $-$1.81 & 0.02 &   $-$4.68 &   $-$1.27 & 0.02 \\
Y\,II  &   1 & $< -$1.36 & 0.18 & $< -$3.57 & $< -$0.16 & 0.18 \\
Zr\,II &   1 & $<$  0.08 & 0.18 & $< -$2.50 & $< +$0.91 & 0.18 \\
Ba\,II &   1 & $< -$2.28 & 0.16 & $< -$4.46 & $< -$1.05 & 0.16 \\
La\,II &   1 & $< -$1.08 & 0.17 & $< -$2.18 & $< +$1.23 & 0.17 \\
Eu\,II &   1 & $< -$1.73 & 0.16 & $< -$2.25 & $< +$1.16 & 0.16 \\
\enddata
\tablenotetext{a}{Applying a $-$0.6 dex correction to the 1D abundance
\citep{asplund_araa}.}
\tablenotetext{b}{Using the EW measures of \citet{fabbian2009}.}
\tablenotetext{c}{Applying a $-$0.26 dex correction to the LTE
  abundance as calculated by \citet{fabbian2009}, assuming the
{\it S$_{\rm H}$}=1 scaling of
collisions with neutral H atoms.}
\tablenotetext{d}{Applying a $-$0.45 dex correction to the LTE
  abundance as calculated by
\citet{fabbian2009}, assuming the {\it S$_{\rm H}$}=1 scaling of
collisions with neutral H atoms.}
\end{deluxetable}

\clearpage

\begin{deluxetable}{lccrccc}
\tablewidth{0pt}
\tabletypesize{\scriptsize}
\tablecaption{\label{tab_nlte_abund} Element Abundances for
  CD~$-$24$^{o}$17504 Based on ``NLTE'' Stellar Parameters}
\tablehead{
\colhead{Species} & \colhead{\#\ lines} & \colhead{log$\epsilon$(X)} &
\colhead{$\sigma$} & \colhead{$\rm[X/H]$} & \colhead{$\rm[X/Fe]$} &
\colhead{$\sigma$/$\sqrt(N)$}
}
\startdata
Li\,I  &   1 &      1.99 & 0.10 &   \nodata &   \nodata & \nodata \\
1D CH  &   2 &      5.98 & 0.04 &   $-$2.46 &   $+$0.83 & 0.03 \\
3D CH\tablenotemark{a}  &   2 & 5.38 & 0.04 & $-$3.05 & $+$0.24 & 0.03\\
LTE C\,I\tablenotemark{b} & 2 & 5.83 & 0.05 & $-$2.60 & $+$0.69 & 0.03\\
NLTE C\,I\tablenotemark{c} & 2& 5.52 & 0.05 & $-$2.91 & $+$0.38 & 0.03\\
LTE O\,I\tablenotemark{b} & 1 & 6.25 & \nodata & $-$2.44 & $+$0.85 & \nodata\\
NLTE O\,I\tablenotemark{d} & 1 & 5.91 & \nodata & $-$2.77 & $+$0.51 & \nodata\\
Na\,I  &   2 &      2.60 & 0.02 &   $-$3.64 &   $-$0.35 & 0.01 \\
Mg\,I  &   6 &      4.53 & 0.04 &   $-$3.07 &   $+$0.22 & 0.02 \\
Al\,I  &   2 &      2.44 & 0.03 &   $-$4.01 &   $-$0.72 & 0.02 \\
Si\,I  &   1 &      4.25 & 0.15 &   $-$3.26 &   $+$0.03 & 0.15 \\
Ca\,I  &  11 &      3.18 & 0.09 &   $-$3.16 &   $+$0.13 & 0.03 \\
Sc\,II &   5 &      0.05 & 0.10 &   $-$3.10 &   $+$0.19 & 0.04 \\
Ti\,I  &   4 &      2.20 & 0.05 &   $-$2.75 &   $+$0.54 & 0.03 \\
Ti\,II &  16 &      2.01 & 0.07 &   $-$2.94 &   $+$0.35 & 0.02 \\
Cr\,I  &   5 &      2.24 & 0.08 &   $-$3.40 &   $-$0.11 & 0.04 \\
Mn\,I  &   3 &      2.03 & 0.03 &   $-$3.40 &   $-$0.11 & 0.02 \\
Fe\,I  & 113 &      4.21 & 0.09 &   $-$3.29 &   \nodata & 0.01 \\
Fe\,II &   5 &      4.21 & 0.07 &   $-$3.29 &   \nodata & 0.03 \\
Co\,I  &   4 &      2.16 & 0.11 &   $-$2.83 &   $+$0.46 & 0.06 \\
Ni\,I  &   4 &      3.17 & 0.05 &   $-$3.04 &   $+$0.25 & 0.03 \\
Zn\,I  &   1 &      2.19 & 0.15 &   $-$2.37 &   $+$0.92 & 0.15 \\
Sr\,II &   2 &   $-$1.72 & 0.03 &   $-$4.59 &   $-$1.30 & 0.02 \\
Y\,II  &   1 & $< -$1.25 & 0.18 & $< -$3.46 & $< -$0.17 & 0.18 \\
Zr\,II &   1 & $<$  0.19 & 0.18 & $< -$2.39 & $< +$0.90 & 0.18 \\
Ba\,II &   1 & $< -$2.18 & 0.16 & $< -$4.36 & $< -$1.07 & 0.16 \\
La\,II &   1 & $< -$0.96 & 0.17 & $< -$2.06 & $< +$1.23 & 0.17 \\
Eu\,II &   1 & $< -$1.62 & 0.16 & $< -$2.14 & $< +$1.15 & 0.16 \\
\enddata
\tablenotetext{a}{Applying a $-$0.6 dex correction to the 1D abundance
\citep{asplund_araa}.}
\tablenotetext{b}{Using the EW measures of \citet{fabbian2009}.}
\tablenotetext{c}{Applying a $-$0.31 dex correction to the LTE
  abundance as calculated by \citet{fabbian2009}, assuming the 
{\it S$_{\rm H}$}=1 scaling of
collisions with neutral H atoms.}
\tablenotetext{d}{Applying a $-$0.34 dex correction to the LTE
  abundance as calculated by
\citet{fabbian2009}, assuming the {\it S$_{\rm H}$}=1 scaling of
collisions with neutral H atoms.}
\tablecomments{Except where stated, the abundances here do not include additional element-specific
NLTE corrections that are qualitatively described in relevant sections
in the paper.}
\end{deluxetable}

\end{document}